\documentclass[epsfig,12pt]{article}
\usepackage{makeidx}
\usepackage{amsmath}
\usepackage{amsfonts}
\usepackage{amssymb}
\usepackage{graphicx}
\usepackage{subfig}
\usepackage{cite}

\input epsf.sty
\makeatletter \@addtoreset{equation}{section}
\renewcommand{\theequation}{\thesection.\arabic{equation}}
\input{tcilatex}
\begin{document}

\title{\textbf{One-loop Type II Seesaw Neutrino Model with Stable Dark
Matter Candidates}}
\date{}
\author{M. A. Loualidi\thanks{%
E-mail: mr.medamin@gmail.com} \textsuperscript{\textsection} and M. Miskaoui%
\thanks{%
E-mail: m.miskaoui@gmail.com} \textsuperscript{\textsection} \\
{\small LPHE-Modeling and Simulations, Faculty of \ Sciences, Mohammed V
University,}\\
\ \ \ \ {\small Rabat, Morocco}}
\maketitle

\begin{abstract}
Opening up the Weinberg operator at 1-loop level using a scalar
triplet, two scalar doublets and one fermion gives rise to T4-2-i
one-loop topology. Neutrino masses generated from this topology are
always accompanied by the tree level Type II seesaw contribution. In
this work, we propose a radiative Majorana neutrino mass model based
on this topology where to avoid the tree
level Type II seesaw mechanism, we extend the model by a $\emph{G}_{f}=%
\mathbb{D}_{4}\times Z_{3}\times Z_{5}\times Z_{2}$ flavor symmetry
and we promote the\ fermion\ inside\ the\ loop\ to three
right-handed neutrinos. In this scenario, the tree level Dirac
neutrino masses resulted from these right-handed neutrinos is also
prevented by the $\emph{G}_{f}$\ group. Moreover, in order for
T4-2-i topology to fully function, the scalar sector is extended by
two flavon fields where after $\emph{G}_{f}$ symmetry breaking, the
model accounts successfully for the observed neutrino masses and
mixing as well as allows for the existence of stable dark matter
(DM) candidates. Indeed, all the particles running in the loop are
potential dark matter candidates as their stability is guaranteed by
the unbroken discrete group\textbf{\ }$Z_{2}\in \emph{G}_{f}$.

\emph{Key words}: Neutrino masses and mixing, Flavor symmetries, Dark matter
stability.
\end{abstract}

\begingroup \renewcommand \thefootnote{\textsection} \footnotetext{%
Both authors contributed equally to this work.} \endgroup

\newpage

\section{Introduction}

The developments in the field of neutrino physics in the past two decades
have been undoubtedly impressive. Neutrinos which rarely interact with
ordinary matter have been identified in the Standard Model (SM) as massless
particles. However, many neutrino oscillation experiments performed in the
past twenty years confirmed that neutrinos have nonzero masses, thus making
these particles as the current best probe for new physics beyond the SM
(BSM) \textrm{\cite{R1,A1}}. Another matter that requires going BSM and
which has been explored at length in the literature is\textrm{\ }the
existence of dark matter where amongst its known properties, an appropriate
candidate must has zero electric charge, produce the correct relic abundance
and must be stable over cosmological time scales \textrm{\cite{R2}}. This
stability asserts the existence of a new kind of charge carried by the DM
particle, and in model building, the stability is usually guaranteed by
imposing new symmetries like $Z_{2}$ which is the most commonly used
symmetry in the literature.\newline
In recent years, there have been a growing interest in radiative neutrino
mass models that provide an interconnection between the neutrino and the DM
sectors. Indeed, these models predict neutrino masses at the loop level as
well as the existence of DM candidates in the form of one of the
intermediate particles running in the loop. One class of these models is the
n-loop realizations of the well-known $d=5$\ Weinberg operator $O_{5}=LLHH$
where $L$ stands for the $SU(2)_{L}$ lepton doublets while $H$ denotes the $%
SU(2)_{L}$ Higgs doublet of the SM\footnote{%
For a systematic investigation of radiative Dirac neutrino mass models
emerging from one-loop and two-loop topologies, see for instance \cite{X1}
and references therein.}. A popular one-loop realization of $O_{5}$\ is the
scotogenic model which extends the SM particle content by three right-handed
neutrinos and an extra inert scalar doublet \cite{R3}, while an exact $Z_{2}$%
\ symmetry prevents the tree level Dirac masses for neutrinos as will as
allowing for stable DM candidates. This model has been studied in detail
using the same and in many times different set of particles inside the loop;
see, for instance, Refs.\textrm{\ \cite%
{R4,R5,A2,A3,R6,A4,R7,A5,R8,A6,R9,R10,R11,R12,R13,R14,R15,R16,R17,R18,A7,R19,R20,R21,R22,R23,R24,R25,R26,R27,R28,R29,R30,R31,R32,R33,R34,R35,R36,R37,R38}%
. }The full possible one-loop diagrams induced from this operator can be
found in \textrm{\cite{R39}} while a systematic study of two and three-loop
realizations of $O_{5}$\ is done in \textrm{\cite{R40,A8}} and \textrm{\cite%
{R41}}, respectively. For a detailed review on radiative neutrino mass
models and their classification see\textrm{\ \cite{R42} }and the references
therein.\textrm{\ }To explain neutrino data along with providing a good DM
candidate in the context of radiative models, the particle content and the
gauge symmetry of the SM, $G_{SM}=SU(3)_{C}\times SU(2)_{L}\times U(1)_{Y}$,
need to be extended. Actually, there are no restrictions concerning whether
the extra symmetries should be Abelian or non-Abelian, discrete or
continuous, simple or multiple. On the other hand, it is well-known that
non-Abelian discrete groups are well justified by the large leptonic mixing
angles measured by the oscillation experiments, and when radiative models
are extended by a non-Abelian flavor symmetry, an interesting implication is
that the stability of DM candidate may be ensured by one of the subgroups
obtained after breaking the flavor symmetry, see for instance Refs. \textrm{%
\cite{R43,R44,R45,R46,R47,R48,R49,R50,R51,R52,R53}}. Therefore, non-Abelian
flavor symmetries are an effective tool to address both neutrino and dark
matter issues.

While most of the finite one-loop diagrams are studied extensively in
model-building BSM, there is in particular one topology that have never been
realized in a field theory; it is denoted by T4-2-i as illustrated in figure %
\ref{01} \textrm{\cite{R16,R39}}. This topology involves a scalar triplet $T$%
---with hypercharge $Y=2$---two extra inert scalars $\phi $\ and $\rho $\
and one fermion $\psi $\ running in the loop. The Higgs triplet $T$\ couples
to the SM Higgs doublet $H$\ through the interaction $HT^{\dagger }H$,\ and
thus, it will always involves the usual tree level Type II seesaw\footnote{%
Dark matter and neutrino mass problems are also sudied in models where
neutrino masses are generated by the tree level TypeII seesaw model
mechanism, see, for instance \cite{A9,A10}.} contribution to neutrino masses
$LTL$\ that cannot be prevented by any additional $U(1)$\ or $Z_{N}$\
symmetries \textrm{\cite{R16,R39}}. The authors in reference \textrm{\cite%
{R39}} stated that to prevent the tree level contributions, two things are
required: $\emph{(i)}$ Promoting the\ fermion\ $\psi $\ inside\ the\ loop\
to\ be\ Majorana\ fermion; and \emph{(ii)} assuming\ that\ all\ couplings\
conserve lepton number.

In this paper,\ our purpose is to cure the difficulties encountered when
building a field theory with topology T4-2-i. To achieve this, we propose a
radiative Majorana neutrino mass model within an extension of the SM based
on the\textrm{\ }$\emph{G}_{f}=\mathbb{D}_{4}\times Z_{3}\times Z_{5}\times
Z_{2}$ flavor symmetry. Furthermore, as previously mentioned, in order to
obtain neutrino masses and mixing consistent with the current neutrino data
along with providing a stable DM candidate, the obvious implication is that
we must extend the particle content of the SM as well. Therefore, we proceed
with the first requirement in \textrm{\cite{R39}} and we promote the\
fermion\ inside\ the\ loop\ to three right-handed neutrino singlets $N_{k}$,
while we discard the second one; which means that we do not assume that\
all\ couplings\ must conserve lepton number. The alternative for the second
requirement---which ensures the suppression\ of the tree level Type II
seesaw contribution to neutrino masses $LTL$---is fulfilled by the choice of
the particle assignment under $Z_{3}\times Z_{5}\in \emph{G}_{f}$. Actually,
our $Z_{3}\times Z_{5}$ charge assignments given in Tables \ref{t2} and \ref%
{t3} prevent the tree level Type II seesaw contribution as well as the tree
level Dirac Yukawa coupling $y_{ij}\overline{L}_{i}\widetilde{H}N_{j}$, and
eventially, the Type I seesaw contribution to neutrino masses. Thus, the
only possibility for neutrino mass generation in our model is at the loop
level in the scotogenic fashion. However, the price to pay with the $Z_{3}$
charge assignments is that the two Yukawa couplings connecting $N_{k}$, $L$
and the two inert scalars in the loop of topology T4-2-i carry non trivial $%
Z_{3}$ charge. Moreover, the usual vertex\ connecting two Higgs doublets
with the scalar triplet $T$ (upper vertex in figure \ref{01}) is also
prevented by the $Z_{3}$\ symmetry. To restor $Z_{3}$ invariance, we have
enlarged the scalar sector by adding two flavon fields $\digamma $ and $\chi
$\ carrying quantum numbers under $\emph{G}_{f}$; thus, fixing the issues of
topology T4-2-i.\textrm{\ }When the flavon $\digamma $ acquires its vacuum
expectation value (VEV), the $\mathbb{D}_{4}$ group gets broken down to a
subgroup $Z_{2}^{\prime }$ leading to a neutrino mass matrix compatible with
the well-known trimaximal mixing matrix \textrm{\cite%
{R54,R55,R56,R57,R58,R59,R60}}. We have studied numerically the
phenomenology associated with neutrino sector in the normal mass hierarchy
(NH) case. Finally, for the DM candidates, all the particles running in the
loop---right-handed neutrino $N_{k}$\ and the scalars $\rho $\ and $\phi $%
---are odd under the discrete group $Z_{2}\in \emph{G}_{f}$ whilst all SM
particles are even. Therefore, the lightest odd particle will be stable and
can play the role of the DM candidate.\emph{\ }We have discussed the
validity of DM candidates for two cases; (a) Fermionic DM candidate with $%
N_{3}$\ being the lightest odd particle, and (b) Bosonic DM candidate with $%
\rho $\ being the lightest odd particle. On the other hand, although the $%
Z_{2}^{\prime }$ subgroup of $\mathbb{D}_{4}$ is unbroken, it is not
responsible for DM stability; however, there might be processes allowed by $%
Z_{2}$\ but forbidden by $Z_{2}^{\prime }$\ since the residual symmetry that
survives the $\emph{G}_{f}$\ symmetry breaking is given by the group $%
Z_{2}^{\prime }\times Z_{2}$. Thus, we have checked the invariance of the
various DM processes under $Z_{2}$\ as well as $Z_{2}^{\prime }$.

The paper is organized as follows. In Sec. II we start by a general
discussion on topology T4-2-i, then we present our field content and the
solution to the problems of topology T4-2-i. In Sec. III we start by
studying in details the neutrino sector and then describe the phenomenology
associated with neutrino masses and mixing.\emph{\ }In Sec. IV we discuss
the dark matter sector where we comment briefly the cases of fermionic and
bosonic DM candidates\emph{.} In Sec. V, we give our conclusion. Finally, we
add an Appendix which contains some useful tools on the dihedral\textrm{\ }$%
\mathbb{D}_{4}$ group.

\section{Genuine one-loop Type II seesaw using $\emph{G}_{f}$ flavour
symmetry}

In this section, we first describe the particles involved in topology T4-2-i
and all their possible charge assignments under the electroweak (EW) gauge
group and we provide the necessary requirements to fix the issues associated
with topology T4-2-i. Then, we present our scenario to account for this
topology by implementing the $\emph{G}_{f}$\ flavor symmetry accompanied
with extra flavon fields.

\subsection{One loop Type II seesaw topology}

\begin{figure}[th]
\begin{center}
\includegraphics[width=.34\textwidth]{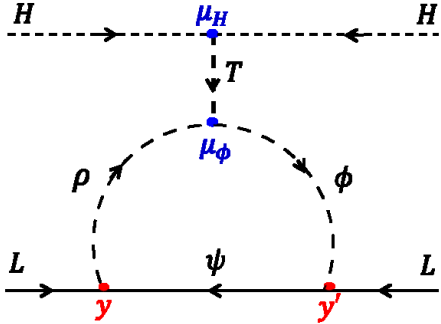}\quad
\end{center}
\par
\vspace{-0.6cm}
\caption{One-loop neutrino mass generation from an $SU(2)_{L}$ scalar
triplet like in the Type II seesaw mechanism. This diagram is denoted as
T4-2-i in reference \protect\cite{R39}.}
\label{01}
\end{figure}

There are several approaches to generate neutrino masses beyond the SM,
among which are the radiative models where neutrino masses arise at the loop
level. These models are rather interesting because they not only account for
the tiny neutrino masses naturally, but also provide a DM candidate given by
one of the new fields running in the loop. One of the most effective ways to
classify these models is through the topology of the loop diagrams which
generate neutrino masses \textrm{\cite{R39,R40,R42,R61,R62}}. The majority
of these models are the one-loop realizations of the well-known dimension-5
Weinberg operator $LLHH$. While most of the finite one-loop diagrams are
studied extensively in building BSM physics models, there is in particular
one topology that have never been realized in a field theory; it is denoted
by T4-2-i as illustrated in Fig. \ref{01}.
In this topology, there are four new particles compared to the SM; an $%
SU(2)_{L}$ scalar triplet with hypercharge $Y=2$ which couples to the SM
Higgs doublets $H$ (bottom vertex), two scalars $\phi $\ and $\rho $\ and
one fermion $\psi $\ running in the loop. From this, we deduce five
different field assignments leading to five different models generating
neutrino masses at one-loop. These five\textrm{\ }possibilities are reported
in Table\textrm{\ }\ref{t1}\textrm{\ }using $SU(2)_{L}$ representations to
differentiate between different models
\begin{table}[h]
\centering%
\begin{tabular}{|l||l|l|l|l|l|}
\hline
$\text{Fields}$ & $\text{Model I}$ & $\text{Model II}$ & $\text{Model III}$
& $\text{Model IV}$ & $\text{Model V}$ \\ \hline
$\phi $ & $\mathbf{3}$ & $\mathbf{2}$ & $\mathbf{2}$ & $\mathbf{1}$ & $%
\mathbf{3}$ \\ \hline
$\rho $ & $\mathbf{1}$ & $\mathbf{2}$ & $\mathbf{2}$ & $\mathbf{3}$ & $%
\mathbf{3}$ \\ \hline
$\psi $ & $\mathbf{2}$ & $\mathbf{1}$ & $\mathbf{3}$ & $\mathbf{2}$ & $%
\mathbf{2}$ \\ \hline
\end{tabular}%
\caption{Different $SU(2)_{L}$ assignments for the fields $\protect\rho $, $%
\protect\phi $ and $\protect\psi $ leading to five possible one-loop
neutrino mass models from topology T4-2-i.}
\label{t1}
\end{table}

On the other hand, it was mentioned in Refs. \textrm{\cite{R16,R39}} that
topology T4-2-i will always involves the usual tree level Type II seesaw
contribution to neutrino masses $LTL$ that cannot be prevented by any
additional $U(1)$ or $Z_{N}$ symmetries. This can be easily shown by
considering the hypercharge quantum numbers of the different particles
involved in the tree level contribution as well as topology T4-2-i.
Therefore, for the Type II seesaw mass term $LTL$ we have the condition%
\begin{equation}
2Y_{L}+Y_{T}=0\quad \text{with}\quad Y_{L}=-1\text{ and }Y_{T}=2,  \label{e1}
\end{equation}%
where $Y_{X}$ is the hypercharge of field $X$ under the $U(1)_{Y}$ group.
For topology T4-2-i, the loop in the diagram of Fig. \ref{01} consists of
three vertices with the following conditions on $Y_{X}$%
\begin{equation}
\begin{array}{ccc}
Y_{L}-Y_{\rho }+Y_{\psi }=0 & \rightarrow & \text{vertex connecting }L\text{%
, }\phi \text{ and }\psi \\
Y_{L}+Y_{\phi }-Y_{\psi }=0 & \rightarrow & \text{vertex connecting }L\text{%
, }\rho \text{ and }\psi \\
Y_{T}+Y_{\rho }-Y_{\phi }=0 & \rightarrow & \text{vertex connecting }T\text{%
, }\phi \text{ and }\rho .%
\end{array}%
\end{equation}%
The sum of these three equations leads to the condition (\ref{e1}) which
implies that a neutrino mass generated by topology T4-2-i is always
accompanied by the tree-level Type II seesaw mechanism. This is true for any
$U(1)$ or $Z_{n}$ quantum charges $q_{X}$. On the other hand, the authors in
Refs. \textrm{\cite{R16,R39}} stated that to prevent the tree level
contributions, two things are required: \emph{(i)} Promoting the\ fermion\ $%
\psi $\ inside\ the\ loop\ to\ be\ Majorana\ fermion; and \emph{(ii)}
assuming\ that\ all\ couplings\ conserve lepton number. In this regard, once
these two conditions are imposed, the tree level Type II seesaw contribution
$LTL$ will be eliminated as it violates lepton number conservation while the
Majorana\ mass term for the fermion running in the loop $M_{i}\bar{\psi}%
_{i}^{c}\psi _{i}$\ will be the only term allowed to break lepton number.
Moreover, these two conditions narrow down the\ number\ of\ $SU(2)_{L}\times
U(1)_{Y}$ assignments\ for the\ fermion\ $\psi $ to\ only\ two\ options:\ a
fermion singlet or a fermion triplet both with hypercharge $Y=0$. As a
result, only the assignments in the models II and III from Table \ref{t1}
are allowed in this scenario. However, building models and taking into
account these prerequisites---especially the condition of imposing lepton
number conservation---is not an easy task; thus, a call for additional
symmetries and particles seems necessary\textrm{. }In this regard, we
propose in the next subsection a solution to the issues of topology T4-2-i
by extending the SM by a $\mathbb{D}_{4}\times Z_{3}\times Z_{5}\times Z_{2}$
flavor symmetry\textrm{.}

\subsection{Implementing $\emph{G}_{f}$ flavour symmetry in T4-2-i model}

As mentioned above, the first step to forbid the tree level Type II seesaw
coupling $\lambda LTL$ is by promoting the fermion $\psi $ inside the loop
to a Majorana fermion. In this work, we consider three right-handed neutrino
singlets $N_{k}$ which correspond to model II in Table (\ref{t1}). In a
second step, we extend the SM gauge group with an additional $\emph{G}_{f}=%
\mathbb{D}_{4}\times Z_{3}\times Z_{5}\times Z_{2}$\ flavor symmetry along
with extra flavon fields allowing us to control the couplings in the 1-loop
diagram. Actually, the choice of this additional symmetry in our model is
introduced not only to forbid the tree level Type II seesaw contribution,
but also to satisfy the following requirements: \emph{(i)} forbid the tree
level Type I seesaw contribution coming from the Dirac operator $y_{ij}%
\overline{L}_{i}\widetilde{H}N_{j}$; \emph{(ii)} obtain neutrino masses and
mixing angles consistent with the current neutrino data; and \emph{(iii)}
stabilize the dark matter candidate against decay.
\begin{figure}[th]
\begin{center}
\includegraphics[width=.38\textwidth]{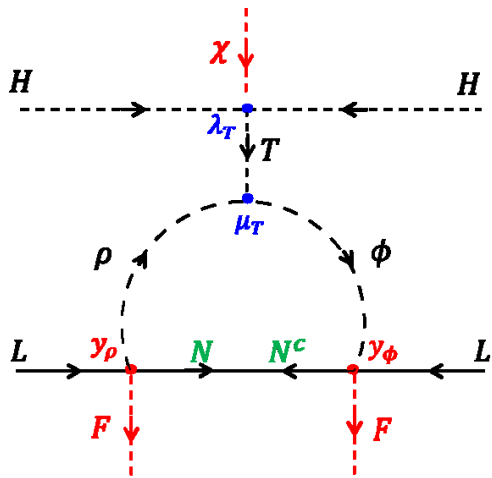}
\end{center}
\par
\vspace{-0.6cm}
\caption{One-loop feynman diagram responsible for the neutrino mass matrix
in our $\mathbb{D}_{4}\times Z_{3}\times Z_{5}\times Z_{2}$ model.}
\label{02}
\end{figure}
Now we turn to present the particle content of the model and describe the $%
\emph{G}_{f}$ quantum numbers of the leptons as well as for new extra
fields. Recall first that the discrete $\mathbb{D}_{4}$ group has five \text{%
irreducible\ representations}: four singlets $1_{p,q}$ with indices $p,q=\pm
,\pm $; and one doublet $2$ indexed by the characters $\mathbf{\chi }(S)$, $%
\mathbf{\chi }$($T$) of the two non-commuting generators $S$ and $T$ of the
dihedral $\mathbb{D}_{4}$; see appendix for more details. For the lepton
sector, as stated in the beginning of this subsection we have added three
right-handed neutrinos to the usual $SU(2)$\ lepton doublets $L_{i}$\ and
lepton singlets $l_{i}^{R}$\ of the SM, here $i$\ run over the three lepton
families. Their quantum numbers under the SM gauge group and the\textrm{\ }$%
\emph{G}_{f}$\textrm{\ }flavor group are as given in Table \ref{t2}.
\begin{table}[h]
\centering%
\begin{equation*}
\begin{tabular}{|l||lll|lll|lll|}
\hline
Fermions & $L_{e}$ & \multicolumn{1}{|l}{$\ \ L_{\mu }$} &
\multicolumn{1}{|l|}{$L_{\tau }$} & $l_{e}^{R}$ & \multicolumn{1}{|l}{$\ \
l_{\mu }^{R}$} & \multicolumn{1}{|l|}{$l_{\tau }^{R}$} & $N_{1}$ &
\multicolumn{1}{|l}{$N_{2}$} & \multicolumn{1}{|l|}{$N_{3}$} \\ \hline\hline
$G_{SM}$ &  & $\left( 1,2\right) _{-1}$ &  &  & $\left( 1,1\right) _{-2}$ &
&  & $\left( 1,1\right) _{0}$ &  \\ \hline
$\mathbb{D}_{4}$ & $1_{+,+}$ & \multicolumn{1}{|l}{\ $\ \ 1_{+,-}$} &
\multicolumn{1}{|l|}{$1_{-,+}$} & $1_{+,+}$ & \multicolumn{1}{|l}{$\ \
1_{+,-}$} & \multicolumn{1}{|l|}{$1_{-,+}$} & $1_{+,+}$ &
\multicolumn{1}{|l}{$\ \ 1_{+,-}$} & \multicolumn{1}{|l|}{$1_{-,+}$} \\
\hline
$\left( Z_{3},Z_{2}\right) $ &  & \ $\ \ \ \left( 1,1\right) $ &  &  & $%
\left( \omega ,1\right) $ &  &  & $\left( 1,-1\right) $ &  \\ \hline
$Z_{5}$ &  & $\ \ \ \ 1$ &  &  & $\ \ \ \eta ^{2}$ &  &  & $\ \ \ 1$ &  \\
\hline
\end{tabular}%
\end{equation*}%
\caption{Gauge and flavor quantum numbers for leptons and right-handed
neutrino fields, where $\protect\omega =e^{\frac{2\protect\pi i}{3}}$ and $%
\protect\eta =e^{\frac{2\protect\pi i}{5}}$.}
\label{t2}
\end{table}
For the scalar sector, besides the usual SM Higgs doublet $H=(h^{+},h^{0})$,
the model involves five additional scalar fields; two inert Higgs doublets $%
\rho =(\rho ^{0},\rho ^{-})$ and $\phi =(\phi ^{+},\phi ^{0})$, one $SU(2)$
scalar triplet $T$ and two flavon fields $\digamma $\ and $\chi $.
\begin{table}[h]
\centering%
\begin{equation*}
\begin{tabular}{|l||l|l|l|l|l|l|}
\hline
Scalars & $H$ & $T$ & $\rho $ & $\phi $ & $\digamma $ & $\chi $ \\
\hline\hline
$G_{SM}$ & $\left( 1,2\right) _{1}$ & $\left( 1,3\right) _{2}$ & $(1,2)_{-1}$
& $(1,2)_{1}$ & $\left( 1,1\right) _{0}$ & $\left( 1,1\right) _{0}$ \\ \hline
$\mathbb{D}_{4}$ & $\ \ \ 1_{+,+}$ & \ $\ 1_{+,+}$ & \ $\ \ \ 2$ & $\ \ \ \
\ 2$ & \ $\ \ \ 2$ & $\ 1_{+,+}$ \\ \hline
$\left( Z_{3},Z_{2}\right) $ & \ $\ \ \left( \omega ^{2},1\right) $ & $%
\left( \omega ^{2},1\right) $ & $\left( \omega ^{2},-1\right) $ & $\left(
\omega ,-1\right) $ & $\left( \omega ,\ 1\right) $ & $\left( \omega
,1\right) $ \\ \hline
$Z_{5}$ & $\ \ \eta ^{3}$ & $\ \ 1$ & $\ \ \ \ \ 1$ & $\ \ \ 1$ & $\ \ \ 1$
& $\ \ \eta ^{4}$ \\ \hline
\end{tabular}%
\end{equation*}%
\caption{Gauge and flavor quantum numbers for all scalar fields of the
model. }
\label{t3}
\end{table}
In our model, the extra right-handed neutrinos $N_{k}$\ and the inert Higgs
doublets $\rho $\ and $\phi $\ are running in the loop as in the original
topology of Fig. \ref{01}. However, based on the $\emph{G}_{f}$\ charge
assignments shown in Table \ref{t3}, the two Yukawa couplings $y_{ik}\bar{L}%
_{i}\rho N_{k}$\ and $y_{jk}^{\prime }\bar{L}_{i}\tilde{\phi}N_{k}$ behave
as doublets under\textrm{\ }$\mathbb{D}_{4}$ group and they carry non zero $%
Z_{3}$ charge $\omega ^{2}$. To restor the invariance under the\textrm{\ }$%
\mathbb{D}_{4}\times Z_{3}$ symmetry, we have added the flavon field $%
\digamma $ which transforms as a $\mathbb{D}_{4}$ doublet and carries a $%
Z_{3}$ charge $\omega $. On the other hand, the one-loop vertex $\mu
_{H}HT^{\dagger }H$\ connecting two Higgs doublets with the scalar triplet
in Fig. \ref{01}\ is prevented in our model by the $Z_{3}$\ symmetry since
its charge is $\omega ^{2}$, the invariance is restored by the flavon field $%
\chi $\ which carries the charge $\omega $, see Table \ref{t3}. The resulted
couplings are invariant under the $Z_{2}$\ symmetry which will be only used
to stabilize the dark matter candidate. Moreover, to guarantee a genuine
1-loop neutrino mass model---no tree level contribution to neutrino
masses---, the dimension-5 operator $L_{i}^{T}TL_{j}\chi $\ which is allowed
by the\textrm{\ }$\mathbb{D}_{4}\times Z_{3}$\textrm{\ }symmetry and leads
to a Type II seesaw\ tree-level contribution is prevented by the discrete $%
Z_{5}\in G_{f}$\ symmetry under which this terms transform as $\eta ^{4}$;
see Tables \ref{t2} and \ref{t3} for the $Z_{5}$\ quantum numbers\ of the
matter and scalar fields respectively.\textbf{\ }Therefore, the $\mathbb{D}%
_{4}\times Z_{3}\times Z_{5}$ group and the new flavon fields are sufficient
to address the challenge of Topology T4-2-i, leading subsequently to the
modified one-loop radiative diagram shown in Fig. \ref{02}. In the following
section, we will study in details the neutrino masses and mixing and their
corresponding phenomenological consequences.

Before we describe the neutrino sector, let us comment briefly on the
charged lepton masses. With respect to the chosen $\mathbb{D}_{4}$\ particle
assignments---see Tables \ref{t2} and \ref{t3}---the charged lepton mass
matrix is diagonal.\textrm{\ }This can easily be seen by considering the
leading order terms responsible for the charged lepton masses. These terms
invariant under $\emph{G}_{f}$ are $y_{e}\left( \bar{L}_{e}\right)
_{++}\left( e_{R}\right) _{++}\left( H\right) _{++}$, $y_{\mu }\left( \bar{L}%
_{\mu }\right) _{+-}\left( \mu _{R}\right) _{+-}\left( H\right) _{++}$ and $%
y_{\tau }\left( \bar{L}_{\tau }\right) _{-+}\left( \tau _{R}\right)
_{-+}\left( H\right) _{++}$. Therefore, after the Higgs field takes its VEV
as $\left\langle H\right\rangle =\left(
\begin{array}{cc}
0 & \frac{1}{\sqrt{2}}\left( \upsilon _{H}+h+iA\right)%
\end{array}%
\right) ^{T}$, we obtain a diagonal charged lepton mass matrix as $%
m_{l}=\upsilon _{H}/\sqrt{2}\mathrm{diag}(y_{e},y_{\mu },y_{\tau })$.
However, it is clear that it is not trivial to produce the mass hierarchy
among charged leptons at the leading order where we\textrm{\ }need to impose
a hierarchical values on the Yukawa couplings, which is considered very
unnatural. On the other hand, in flavor symmetries based models, the mass
hierarchy can be achieved by taking into account corrections in the the
charged lepton mass matrix from higher-dimensional operators involving
flavon fields. An example of such operators can be written as $\bar{L}%
_{l}^{i}l_{R}^{j}H(\frac{\Omega }{\Lambda })^{n}(\frac{\zeta }{\Lambda }%
)^{m} $\ with $n+m\geq 1$ and $\Lambda $\ is a cutoff scale while $\Omega $
and $\zeta $ denote the flavon fields needed also to ensure the invariance
under $\emph{G}_{f}$. Another attractive method used to explain this
hierarchy is the Froggatt-Nielsen mechanism which relies on the spontaneous
breaking of a $U(1)_{F}$ flavor symmetry, for details on this method see
Ref. \textrm{\cite{R63}}.

\section{Neutrino model building based on topology T4-2-i}

In this section, we study the neutrino masses and mixing in the framework
described in the previous subsection. Neutrino masses are generated
radiatively while we considered the trimaximal mixing matrix scheme. Then,
by using the 3$\sigma $\ experimental values of the oscillation parameters,
we show by means of scatter plots the physical observables $m_{ee}$\ and $%
m_{\nu _{e}}$\ related respectively to neutrinoless double beta decay and
tritium beta decay experiments, and we also provide scatter plot predictions
on the sum of neutrino masses as well as on the Dirac $CP$ violating phase.

\subsection{Neutrino masses and mixing}

\begin{figure}[th]
\begin{center}
\hspace{2.5em} \includegraphics[width=.44\textwidth]{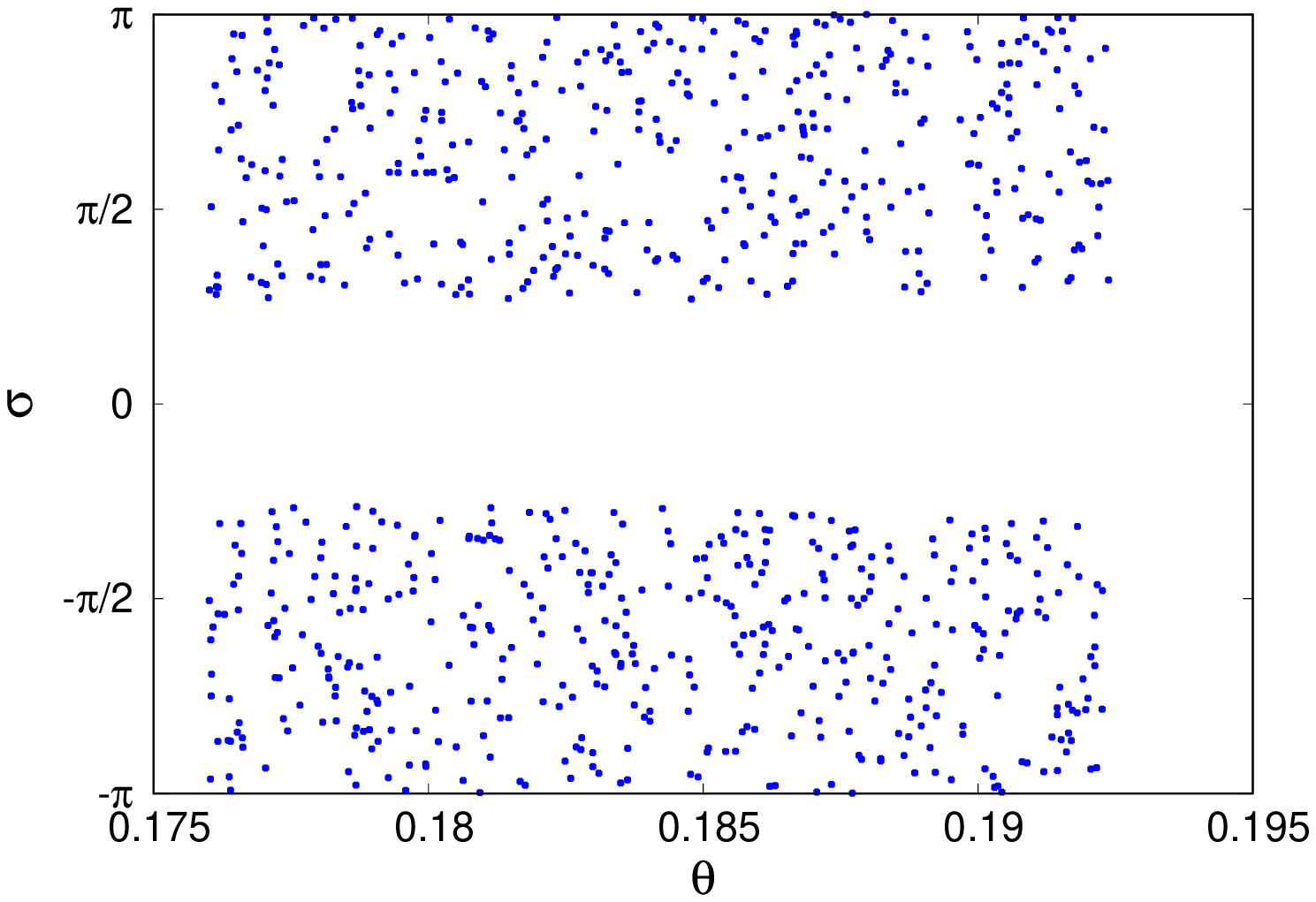}\quad %
\includegraphics[width=.44\textwidth]{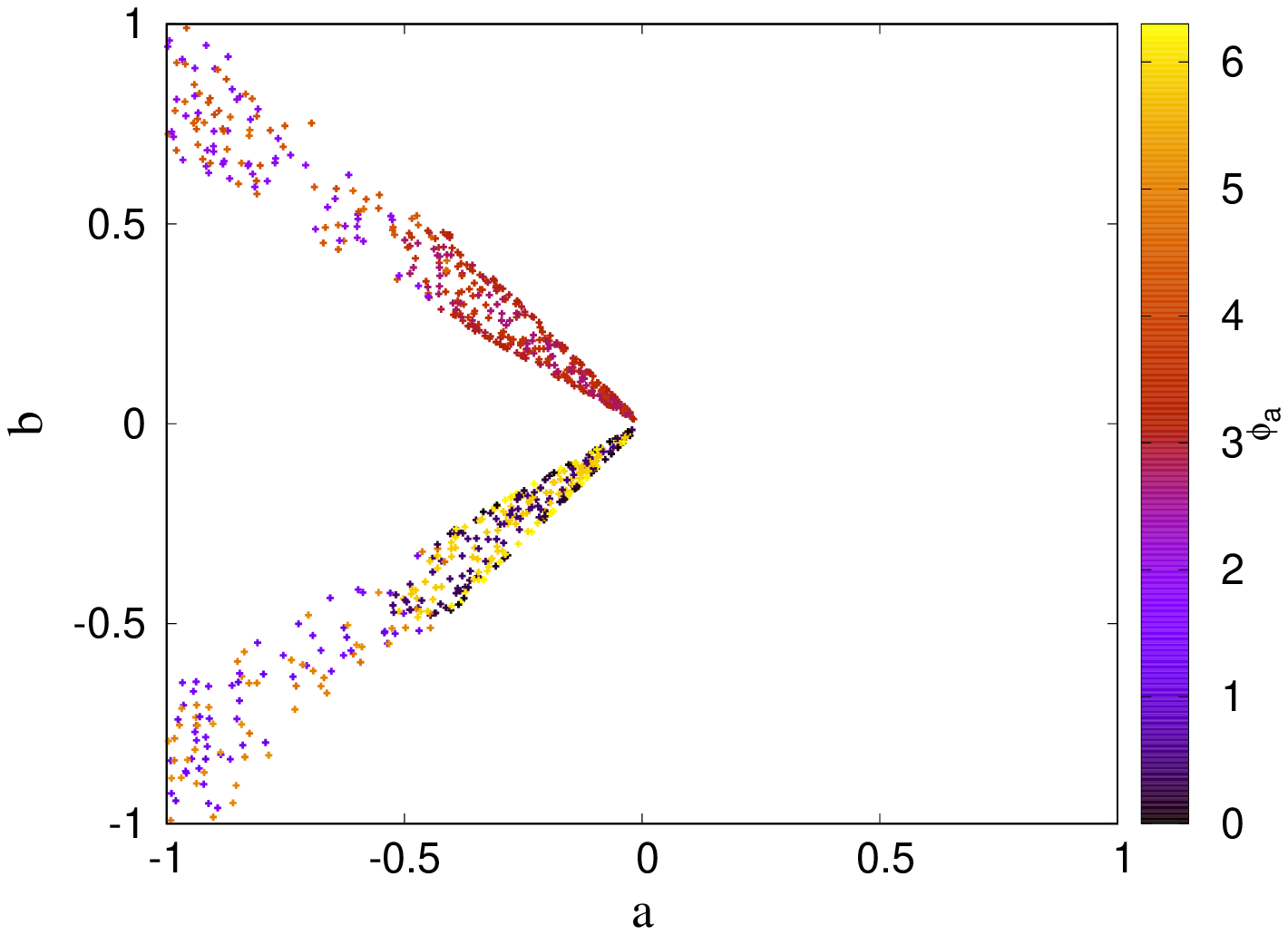} \includegraphics[width=.44%
\textwidth]{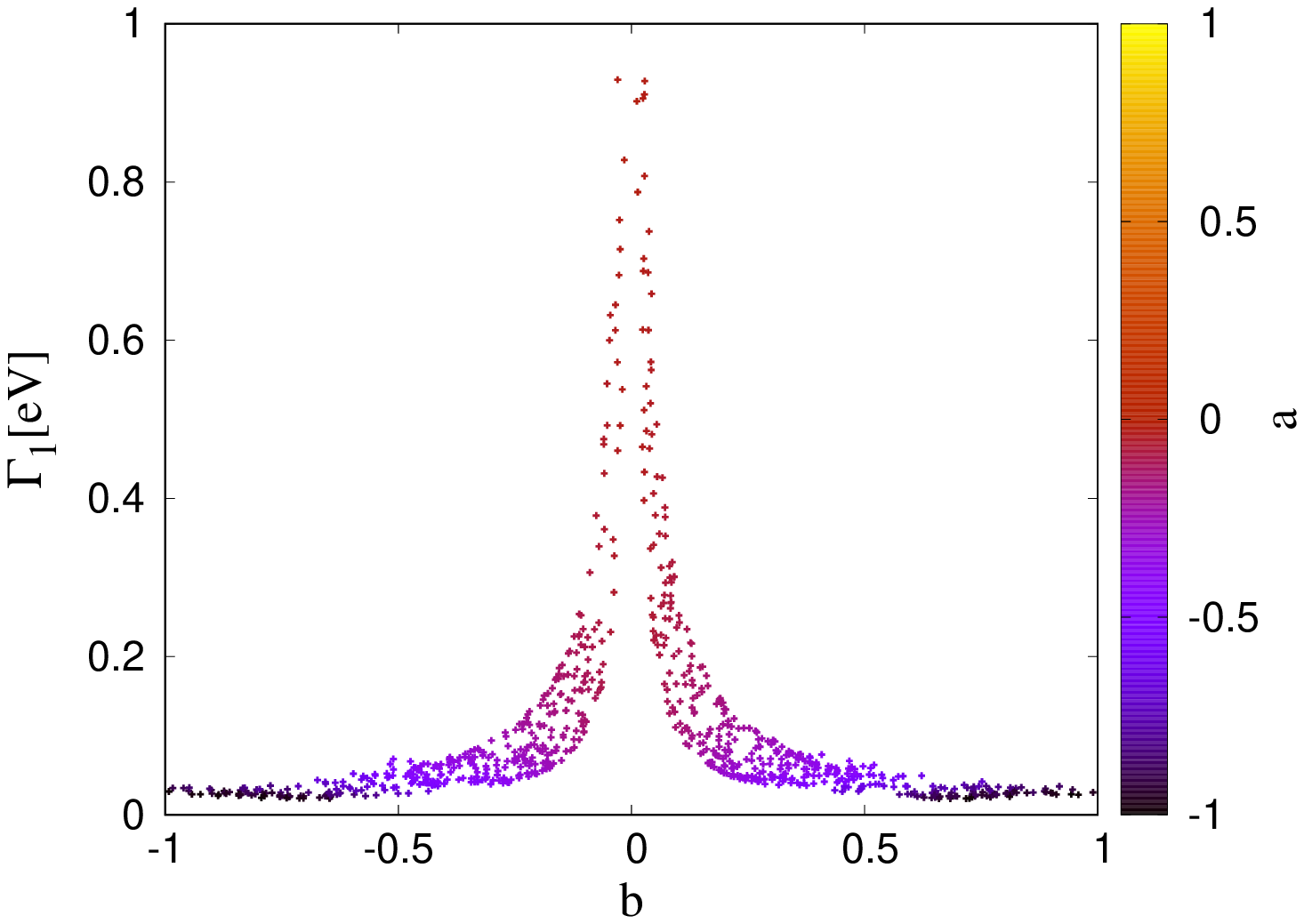}
\end{center}
\par
\vspace{0.5cm}
\caption{Left: Variation of the arbitrary phase $\protect\sigma $ as a
function of the angle $\protect\theta $. Right: Correlation among the
parameters $a$, $b$ and the phase $\protect\phi _{a}$. Bottom: Correlation
among the parameters $a$, $b$ and $\Gamma _{1}$.}
\label{03}
\end{figure}

In our model, the $\emph{G}_{f}$ flavor symmetry forbids the usual SM tree
level Dirac term $y\bar{L}_{i}\widetilde{H}N_{k}$, and since the neutral
component of the scalar fields $\rho $\ and $\phi $ do not acquire VEVs, the
usual seesaw mechanism is no longer responsible for neutrino masses.
Nonetheless, the light neutrino masses are generated radiatively through the
one-loop diagram which involves $\rho $, $\phi $\ and $N_{k}$ in the
internal lines; see Fig. \ref{02}. According to the field assignments in
Tables \ref{t2} and \ref{t3}, the relevant couplings in the neutrino sector,
invariant under gauge and $\mathbb{D}_{4}\times Z_{3}\times Z_{5}\times Z_{2}
$ symmetries are given by the following lagrangian%
\begin{equation}
\mathcal{L}=\frac{y_{\rho }^{ik}}{\Lambda }\bar{L}_{i}N_{k}\rho \digamma +%
\frac{y_{\phi }^{jk}}{\Lambda }\bar{L}_{j}N_{k}\tilde{\phi}\digamma +\frac{%
M_{k}}{2}\overline{N_{k}^{c}}N_{k}+h.c.,  \label{3-1}
\end{equation}%
Here $y_{\rho }^{ik}$ and $y_{\phi }^{jk}$\ are Yukawa couplings and $\tilde{%
\phi}=i\sigma _{2}\phi ^{\ast }$. The first two terms in this lagrangian are
the leading order contributions to Dirac neutrino masses while the third one
is the Majorana mass term for $N_{k}$. For example, the first coupling
transforms under the\textrm{\ }$\mathbb{D}_{4}$\textrm{\ }discrete symmetry
as%
\begin{equation}
\bar{L}_{i}N_{k}\rho \digamma \sim 1_{a,b}\otimes 1_{c,d}\otimes 2\otimes 2,
\label{3-2}
\end{equation}%
with indices $a,b,c,d=\pm $. Thus, to obtain the desired $\mathbb{D}_{4}$
trivial singlet, the tensor product between the $\mathbb{D}_{4}$
doublets---which decomposes into the direct sum of the four $\mathbb{D}_{4}$
singlets; see the Appendix---should transform in the same manner as the
product between the two singlet $1_{a,b}\otimes 1_{c,d}$. This can easily be
seen in the following examples%
\begin{eqnarray}
\bar{L}_{e}N_{1}\rho \digamma  &\sim &\left. \left( 1_{+,+}\otimes
1_{+,+}\right) \right\vert _{1_{+,+}}\otimes \left. \left( 2\otimes 2\right)
\right\vert _{1_{+,+}}  \notag \\
\bar{L}_{e}N_{2}\rho \digamma  &\sim &\left. \left( 1_{+,+}\otimes
1_{+,-}\right) \right\vert _{1_{+,-}}\otimes \left. \left( 2\otimes 2\right)
\right\vert _{1_{+,-}}.
\end{eqnarray}%
The same discussion holds for the second term in (\ref{3-1}). To break the
flavor symmetry, the flavon doublet $\digamma $ acquires its VEV along the
direction $\left\langle \digamma \right\rangle =\upsilon _{\digamma }\left(
1,1\right) $ while the scalar fields $\rho $\ and $\phi $\ do not acquire
VEVs and may be expressed as%
\begin{equation}
\begin{tabular}{ccc}
$\rho =\left(
\begin{array}{c}
\frac{1}{\sqrt{2}}\left( \rho _{1}+i\rho _{2}\right)  \\
\rho ^{-}%
\end{array}%
\right) $ & , & $\phi =\left(
\begin{array}{c}
\phi ^{+} \\
\frac{1}{\sqrt{2}}\left( \phi _{1}+i\phi _{2}\right)
\end{array}%
\right) ,$%
\end{tabular}
\label{3-3}
\end{equation}%
with $\rho _{1}$\ ($\phi _{1}$) and $\rho _{2}$($\phi _{2}$) present
respectively the scalar and the pseudoscalar parts of the neutral component
of $\rho $\ ($\phi $).\emph{\ }At the first sight, it seems that $\rho $\
and $\phi $\ are adjoint of each other as they carry the hypercharges $Y=-1$%
\ and $Y=+1$\ and $Z_{3}$\ charges $\bar{\omega}$\ and $\omega $\
respectively. However, they transform in the following manner under $\mathbb{%
D}_{4}$%
\begin{equation}
\hat{\rho}=\left(
\begin{array}{c}
\rho  \\
0%
\end{array}%
\right) \quad \text{and}\quad \hat{\phi}=\left(
\begin{array}{c}
\phi  \\
0%
\end{array}%
\right)   \label{3-4}
\end{equation}%
in such a way that $\rho $\ and $\phi ^{\dagger }$\ are placed in different%
\textrm{\ }$\mathbb{D}_{4}$\textrm{\ }doublet components; see Appendix for
more details on $\mathbb{D}_{4}$ group properties. This difference between $%
\rho $\ and\textrm{\ }$\phi ^{\dagger }$ is due to the vertex connecting $T$%
, $\phi $\ and $\rho $\ in the diagram of Fig. \ref{02}, where by asking for
a non-vanishing coupling $\mu _{T}T\rho \phi ^{\dagger }$\ the bilinear term
$(\rho \phi ^{\dagger })$\ must transform as a trivial singlet (since $T\sim
1_{++}$). Using the $\mathbb{D}_{4}$ tensor product, the product between the
two mass matrices deduced from the two first terms in (\ref{3-2}) is given by%
\begin{equation}
\frac{\upsilon _{\digamma }^{2}}{\Lambda ^{2}}y_{\rho }^{ik}y_{\phi }^{jk}=%
\frac{\upsilon _{\digamma }^{2}}{\Lambda ^{2}}\left(
\begin{array}{ccc}
y_{\rho }^{e1} & y_{\rho }^{\mu 1} & y_{\rho }^{\tau 1} \\
y_{\rho }^{e2} & y_{\rho }^{\mu 2} & y_{\rho }^{\tau 2} \\
y_{\rho }^{e3} & y_{\rho }^{\mu 3} & y_{\rho }^{\tau 3}%
\end{array}%
\right) \left(
\begin{array}{ccc}
y_{\phi }^{e1} & y_{\phi }^{e2} & -y_{\phi }^{e3} \\
y_{\phi }^{\mu 1} & y_{\phi }^{\mu 2} & -y_{\phi }^{\mu 3} \\
-y_{\phi }^{\tau 1} & -y_{\phi }^{\tau 2} & y_{\phi }^{\tau 3}%
\end{array}%
\right) .
\end{equation}%
As for the Majorana mass term $M_{k}\overline{N_{k}^{c}}N_{k}$, since the
three right-handed neutrinos transform trivially under $\mathbb{D}_{4}\times
Z_{3}\times Z_{5}$, we obtain a diagonal Majorana neutrino mass matrix%
\textrm{\ }$M_{k}=\mathrm{diag}(M_{1},M_{2},M_{3})$. Consequently, neutrino
masses induced via the one-loop diagram in Fig. \ref{02} are given by%
\begin{eqnarray}
(M_{\nu })_{ij} &=&-\frac{\mu _{T}\lambda _{T}\upsilon _{\chi }\upsilon
_{H}^{2}}{m_{T}^{2}}\frac{\upsilon _{\digamma }^{2}}{\Lambda ^{2}}y_{\rho
}^{ik}M_{k}y_{\phi }^{jk}J(m_{\rho }^{2},m_{\phi }^{2},M_{k}^{2})  \notag \\
&=&\tsum\limits_{k}\frac{\upsilon _{\chi }}{\Lambda }y_{\rho }^{ik}\Gamma
_{k}y_{\phi }^{jk},  \label{nm}
\end{eqnarray}%
where $\upsilon _{\chi }$ is the VEV of the flavon $\chi $ while $\Gamma _{k}
$ is defined as follows
\begin{equation}
\Gamma _{k}=-\frac{\mu _{T}\upsilon _{H}^{2}\lambda _{T}M_{k}}{m_{T}^{2}}%
\frac{\upsilon _{\digamma }^{2}}{\Lambda }J(m_{\rho }^{2},m_{\phi
}^{2},M_{k}^{2}),  \label{ga}
\end{equation}%
while the loop function $J$ is defined as
\begin{eqnarray}
J(m_{\rho }^{2},m_{\phi }^{2},M_{k}^{2}) &=&-\frac{1}{(4\pi )^{2}}\left[
\frac{m_{\rho }^{2}}{(m_{\rho }^{2}-m_{\phi }^{2})(m_{\rho }^{2}-M_{k}^{2})}%
\ln \frac{M_{k}^{2}}{m_{\rho }^{2}}\right.   \notag \\
&&\left. +\frac{m_{\phi }^{2}}{(m_{\phi }^{2}-m_{\rho }^{2})(m_{\phi
}^{2}-M_{k}^{2})}\ln \frac{M_{k}^{2}}{m_{\phi }^{2}}\right] .
\end{eqnarray}%
Assuming for simplicity that we have a quasi-degenerate right-handed
neutrino masses with $M_{3}\simeq M_{2}\simeq M_{1}\ $implying $\Gamma
_{3}\simeq \Gamma _{2}\simeq \Gamma _{1}$. In this case, the total neutrino
mass matrix can be expressed as $M_{\nu }=\Gamma _{1}\left[ \frac{\upsilon
_{\chi }}{\Lambda }y_{\rho }^{ik}y_{\phi }^{jk}\right] $, and by assuming
the following conditions on the Yukawa couplings%
\begin{eqnarray}
y_{\phi }^{\mu 2} &=&y_{\phi }^{\tau 2}=y_{\phi }^{\mu 3}=y_{\phi }^{e3}=0%
\text{ \ \ },\text{ \ }y_{\rho }^{\tau 1}=-y_{\rho }^{\tau 3}\text{\ \ },%
\text{ }y_{\rho }^{e1}=-y_{\rho }^{e3}=-y_{\rho }^{e2}\text{ }  \notag \\
y_{\phi }^{\tau 1} &=&y_{\phi }^{\tau 3}\text{\ \ },\text{ \ \ }y_{\rho
}^{\mu 3}=y_{\rho }^{\mu 2}\text{\ \ },\text{ \ }\frac{y_{\phi }^{\mu 1}}{%
y_{\rho }^{e2}}=-\frac{y_{\phi }^{e1}}{y_{\rho }^{\mu 2}}=\frac{2y_{\phi
}^{e2}}{y_{\rho }^{\mu 1}+y_{\rho }^{\mu 2}}\text{\ \ },\text{ \ }y_{\rho
}^{\tau 2}=\frac{y_{\rho }^{e2}y_{\phi }^{e2}}{y_{\phi }^{\tau 3}},
\end{eqnarray}%
we obtain the total neutrino mass matrix expressed as
\begin{equation}
M_{\nu }=\Gamma _{1}\left(
\begin{array}{ccc}
2a+b & -a & -b \\
-a & a & a \\
-b & a & b%
\end{array}%
\right) ,  \label{mm}
\end{equation}%
where to avoid heavy notations we have introduced the following
parametrization $a=\frac{\upsilon _{\chi }}{\Lambda }y_{\rho }^{e2}y_{\phi
}^{e2}\ $and $b=\frac{\upsilon _{\chi }}{\Lambda }y_{\rho }^{\tau 3}y_{\phi
}^{\tau 3}$. This matrix exhibits the magic symmetry referring to the
equality of the sum of each row and the sum of each column in $M_{\nu }$
\textrm{\cite{R64}}. It is well known that the mass matrix acquiring this
property is diagonalized by the trimaximal mixing matrix $U_{TM_{2}}$ which%
\emph{\ }accounts naturally for the nonzero $\theta _{13}$\ as well as a
possible determination of the $\theta _{23}$\ octant. Therefore, $M_{\nu }$\
is diagonalized as $U_{TM_{2}}^{\dag }M_{\nu }U_{TM_{2}}=\mathrm{diag}%
(m_{1},m_{2},m_{3})$ with $U_{TM_{2}}$ is expressed following the PDG
parametrization for the lepton mixing matrix\textrm{\ }as%
\begin{equation}
U_{TM_{2}}=\left(
\begin{array}{ccc}
\sqrt{\frac{2}{3}}\cos \theta  & \frac{1}{\sqrt{3}} & \sqrt{\frac{2}{3}}\sin
\theta e^{-i\sigma } \\
-\frac{\cos \theta }{\sqrt{6}}-\frac{\sin \theta }{\sqrt{2}}e^{i\sigma } &
\frac{1}{\sqrt{3}} & \frac{\cos \theta }{\sqrt{2}}-\frac{\sin \theta }{\sqrt{%
6}}e^{-i\sigma } \\
-\frac{\cos \theta }{\sqrt{6}}+\frac{\sin \theta }{\sqrt{2}}e^{i\sigma } &
\frac{1}{\sqrt{3}} & -\frac{\cos \theta }{\sqrt{2}}-\frac{\sin \theta }{%
\sqrt{6}}e^{-i\sigma }%
\end{array}%
\right) .U_{P}  \label{3-11}
\end{equation}%
Here $\theta $ is an arbitrary angle that will be related to the observed
neutrino mixing angles $\theta _{ij}$, $\sigma $ is an arbitrary phase that
will be related later on to the Dirac $CP$ phase $\delta _{CP}$ and $U_{P}=%
\mathrm{diag}(1,e^{i\frac{\alpha _{21}}{2}},e^{i\frac{\alpha _{31}}{2}})$ is
a diagonal matrix that encodes the Majorana phases $\alpha _{21}$ and $%
\alpha _{31}$. The Yukawa couplings in the parameters $a$\ and $b$\ are
complex number of order one, hence $M_{\nu }$\ is a complex mass matrix. To
diagonalize $M_{\nu }$\ we take $b$\ to be real without loss of generality
while the parameter $a$\ remains complex; $a\rightarrow \left\vert
a\right\vert e^{i\phi _{a}}$\ where $\phi _{a}$\ is a $CP$ violating phase.
As a result, we obtain the three active light neutrino masses $m_{1}$, $m_{2}
$ and $m_{3}$ expressed explicitly as%
\begin{eqnarray}
\left\vert m_{1}\right\vert  &=&\Gamma _{1}\sqrt{\left\vert a\right\vert
^{2}+4b^{2}+\frac{9\left\vert a\right\vert ^{4}}{4b^{2}}+\left\vert
a\right\vert \left( \frac{3\left\vert a\right\vert ^{2}}{b}+4b\right) \cos
\phi _{a}+6\left\vert a\right\vert ^{2}\cos 2\phi _{a}},  \notag \\
\left\vert m_{2}\right\vert  &=&\Gamma _{1}\left\vert a\right\vert ,
\label{3-12} \\
\left\vert m_{3}\right\vert  &=&\Gamma _{1}\sqrt{\left\vert a\right\vert
^{2}+\frac{9\left\vert a\right\vert ^{4}}{4b^{2}}-\frac{3\left\vert
a\right\vert ^{3}}{b}\cos \phi _{a}},  \notag
\end{eqnarray}%
provided that $\left\vert a\right\vert <\left\vert b\right\vert $ and the
following conditions on $\theta $ and $\sigma $\ hold%
\begin{equation}
\tan 2\theta =-\frac{\sqrt{3\left( \left\vert a\right\vert ^{2}-b^{2}\right)
^{2}+12\left\vert a\right\vert ^{2}b^{2}\sin ^{2}\phi _{a}}}{4\left\vert
a\right\vert b\cos \phi _{a}+3\left\vert a\right\vert ^{2}+b^{2}}\quad
,\quad \tan \sigma =\frac{2\left\vert a\right\vert b\sin \phi _{a}}{%
\left\vert a\right\vert ^{2}-b^{2}}.
\end{equation}%
Regarding the mixing angles, we use the PDG standard parametrization of the
Pontecorvo-Maki-Nakagawa-Sakata (PMNS) matrix \textrm{\cite{R65}}, then we
calculate the three observed neutrino mixing angles in terms of the
trimaximal mixing parameters, we find%
\begin{eqnarray}
\sin ^{2}\theta _{13} &=&\frac{2}{3}\sin ^{2}\theta \quad ,\quad \sin
^{2}\theta _{12}=\frac{1}{3-2\sin ^{2}\theta }  \notag \\
\sin ^{2}\theta _{23} &=&\frac{1}{2}-\frac{3\sin 2\theta }{2\sqrt{3}(3-2\sin
^{2}\theta )}\cos \sigma .  \label{mix}
\end{eqnarray}%
Since the recent experimental data signals that a normal mass ordering is
more preferred than an inverted ordering \textrm{\cite{R66,R67}}, we will
perform our numerical study in the normal hierarchy case. Therefore, we use
as input data the results of the global analysis by NuFIT 4.0 of the
neutrino oscillation parameters at $3\sigma $ interval \textrm{\cite{R67} }%
in the NH case; we have%
\begin{equation}
\begin{array}{c}
\sin ^{2}\theta _{13}\in \left[ 0.02044\rightarrow 0.02437\right] \quad
,\quad \sin ^{2}\theta _{23}\in \left[ 0.428\rightarrow 0.624\right]  \\
\sin ^{2}\theta _{12}\in \left[ 0.275\rightarrow 0.350\right] \quad ,\quad
\frac{\Delta m_{21}^{2}}{10^{-5}}\left[ \mathrm{eV}^{2}\right] \in \left[
6.79\rightarrow 8.01\right]  \\
\frac{\Delta m_{31}^{2}}{10^{-3}}\left[ \mathrm{eV}^{2}\right] \in \left[
2.431\rightarrow 2.622\right] .%
\end{array}
\label{ev}
\end{equation}%
The trimaximal matrix is described by two unknown parameters $\theta $\ and $%
\sigma $\ which are in turn linked to the free parameters $\Gamma _{1}$, $a$%
, $b$\ and $\phi _{a}$\ appearing in the neutrino mass matrix (\ref{mm}).
First, by using the $3\sigma $\ experimental range of $\sin ^{2}\theta _{13}$%
\ and the first equation in (\ref{mix})\ we find the permitted values of $%
\theta $\ as $0.176\lesssim \theta \left[ \mathrm{rad}\right] \lesssim 0.193$%
.\ Inserting this constraint on $\theta $\ in the formula of the solar
mixing angle in (\ref{mix}) allows to restrict the interval of $\theta _{12}$%
\ compared to its $3\sigma $\ allowed range (see Eq. (\ref{ev})) where we
obtain $\sin ^{2}\theta _{12}\in \left[ 0.334\rightarrow 0.341\right] $.%
\emph{\ }Then, by using the experimental values of three mixing angles $\sin
^{2}\theta _{ij}$\textrm{\ }at $3\sigma $ range, we show in the left panel
of Fig. \ref{03} the correlation between $\theta $\ and the arbitrary phase $%
\sigma $\ which is randomly varied in the range $[-\pi \rightarrow \pi ]$.
Accordingly, we find a more constrained range for $\sigma $\ given by%
\begin{equation}
\sigma \left[ \mathrm{rad}\right] \in \lbrack -3.139313\rightarrow
-0.827825]\cup \lbrack 0.846743\rightarrow 3.141149].  \label{sig}
\end{equation}%
On the other hand, since the parameters\textrm{\ }$a=\frac{\upsilon _{\chi }%
}{\Lambda }y_{\rho }^{e2}y_{\phi }^{e2}\ $and $b=\frac{\upsilon _{\chi }}{%
\Lambda }y_{\rho }^{\tau 3}y_{\phi }^{\tau 3}$\textrm{\ }contribute to the
small neutrino masses, the VEV of the flavon $\chi $\ needs to be small and
close to the cutoff scale\textrm{\ }$\upsilon _{\chi }\lesssim $\textrm{\ }$%
\Lambda $\textrm{\ }with $y_{\rho ,\phi }^{e2},y_{\rho ,\phi }^{\tau 3}\sim
\mathcal{O}(1)$; thus, we have allowed $a$\ and $b$\ to vary in the range $%
-1\lesssim a,b\lesssim 1$\ while the phase $\phi _{a}$\ is considered to be
unrestrained;\ $0\lesssim \phi _{a}\lesssim 2\pi $. Based on this, to get an
idea about the order of magnitude of the parameter $\Gamma _{1}$\ using Eq. (%
\ref{ga}), we make the obvious observation using the second equation in Eq. (%
\ref{nm}) that for neutrino masses around the sub-eV order, $\Gamma _{k}$\
should as well vary in the sub-eV range. However, since $\Gamma _{1}$\ will
be used as an input parameter when discussing the neutrino phenomenology, we
need to fix its range.\ To do this, lets get an estimate on the parameters
involved on the righthand side of Eq. (\ref{ga}). Firstly, from Ref. \cite%
{R39} we learn that when the particles running in the loop get their masses
above the EW scale and up to $1TeV$, the loop function $J(m_{\rho
}^{2},m_{\phi }^{2},M_{k}^{2})$\ gets as low as $10^{-9}$GeV$^{-2}$.
Secondly, it is well known in Type II seesaw models that the VEV of the
Higgs triplet $\upsilon _{T}$\ is proportional to\ $\upsilon _{T}\simeq -\mu
\upsilon _{H}^{2}/m_{T}^{2}$. However, in the present model the trilinear
coupling $\mu $\ connecting the Higgs doublets to the scalar triplet is
replaced by the the vertex $\lambda _{T}HT^{\dagger }H\chi $; thus, $\mu $\
is\ equivalent to $\lambda _{T}\upsilon _{\chi }$\ and eventually the
quantity $-\lambda _{T}\upsilon _{H}^{2}/m_{T}^{2}$\ in Eq. (\ref{ga}) turn
out to be approximately proportional to the VEV ratio $\upsilon
_{T}/\upsilon _{\chi }$. Furthermore, the triplet VEV $\upsilon _{T}$ is
constrained by the experimental value of the $\rho $\ parameter $\rho _{%
\mathrm{exp}}=1.00039\pm 0.00019$\ \cite{R65}, which requires $\upsilon
_{T}\lesssim 4GeV$. At this stage, assuming that the flavons VEVs $\upsilon
_{\digamma }$\ and $\upsilon _{\chi }$\ are of the same order and close to
the cutoff scale and\textrm{\ }$\upsilon _{T}=1\mathrm{GeV}$,\ the parameter
$\Gamma _{1}$\ becomes less than or approximately equals to $\mu
_{T}M_{1}\times 10^{-9}$GeV$^{-1}$. Given the suppression factor\textrm{\ }$%
J(m_{\rho }^{2},m_{\phi }^{2},M_{k}^{2})\lesssim 10^{-9}$\textrm{GeV}$^{-2}$,%
\textrm{\ }$M_{1}$\textrm{\ }was assumed to lie at the TeV scale, and thus, $%
\Gamma _{1}\lesssim 10^{-6}\times \mu _{T}$. Finally, the free
dimension-full parameter $\mu _{T}$\ can be chosen to be as small as $0.001%
\mathrm{GeV}$\ to get $\Gamma _{1}\lesssim 1eV$. Gathering all the above
information, we show in the right panel of Fig. \ref{03} the correlation
among the parameters $a$, $b$\ and $\phi _{a}$\ while in the bottom panel of
the same figure we show the correlation among $\Gamma _{1}$, $a$\ and $b$\
where we used as input parameters the $3\sigma $\ allowed ranges of the mass
squared differences $\Delta m_{21}^{2}$\ and $\Delta m_{31}^{2}$\ given in (%
\ref{ev}).\textrm{\ }As a result, we find%
\begin{eqnarray}
a &\in &\left[ -0.99784\rightarrow -0.02587\right] \quad ,\quad b\in \left[
-0.99955\rightarrow 0.98998\right] \ ,\   \notag \\
\phi _{a}\left[ \mathrm{rad}\right]  &\in &\left[ 0.00151\rightarrow 6.28196%
\right] \quad ,\quad \Gamma _{1}\left[ \mathrm{eV}\right] \in \left[
0.02044\rightarrow 0.92926\right] .
\end{eqnarray}

\subsection{Neutrino phenomenology}

Given that neutrino oscillation experiments depend only on the squared-mass
splittings $\Delta m_{21}^{2}$ and $\Delta m_{31}^{2}$, there are three
different approaches employed to determine the absolute scale of neutrino
masses: \emph{(1)} the sum of the three active neutrino masses from
cosmological observations $\sum m_{i}\equiv m_{sum}=m_{1}+m_{2}+m_{3}$,
\emph{(2) }the effective neutrino mass $m_{\nu _{e}}=\left(
\sum_{i}\left\vert U_{ei}\right\vert ^{2}m_{i}^{2}\right) ^{1/2}$ using
kinematic effects\textrm{\ }in beta decay experiments, and \emph{(3) }the
effective Majorana neutrino mass $\left\vert m_{ee}\right\vert =\left\vert
\sum_{i}U_{ei}^{2}m_{i}\right\vert $ in neutrinoless double beta decay ($%
0\nu \beta \beta $) where\ $m_{i}$ are the three neutrino masses and $U_{ei}$
are the elements of the first row of the mixing matrix. In our numerical
study, we use the latest result from the Planck data which when combined
with measurements of the baryon acoustic oscillations (BAO) provides an
upper limit on $m_{sum}$ given by $m_{sum}<0.12\mathrm{eV}$\ at 95\% C.L
\textrm{\cite{R68}}.
\begin{figure}[th]
\begin{center}
\hspace{2.5em} \includegraphics[width=.44\textwidth]{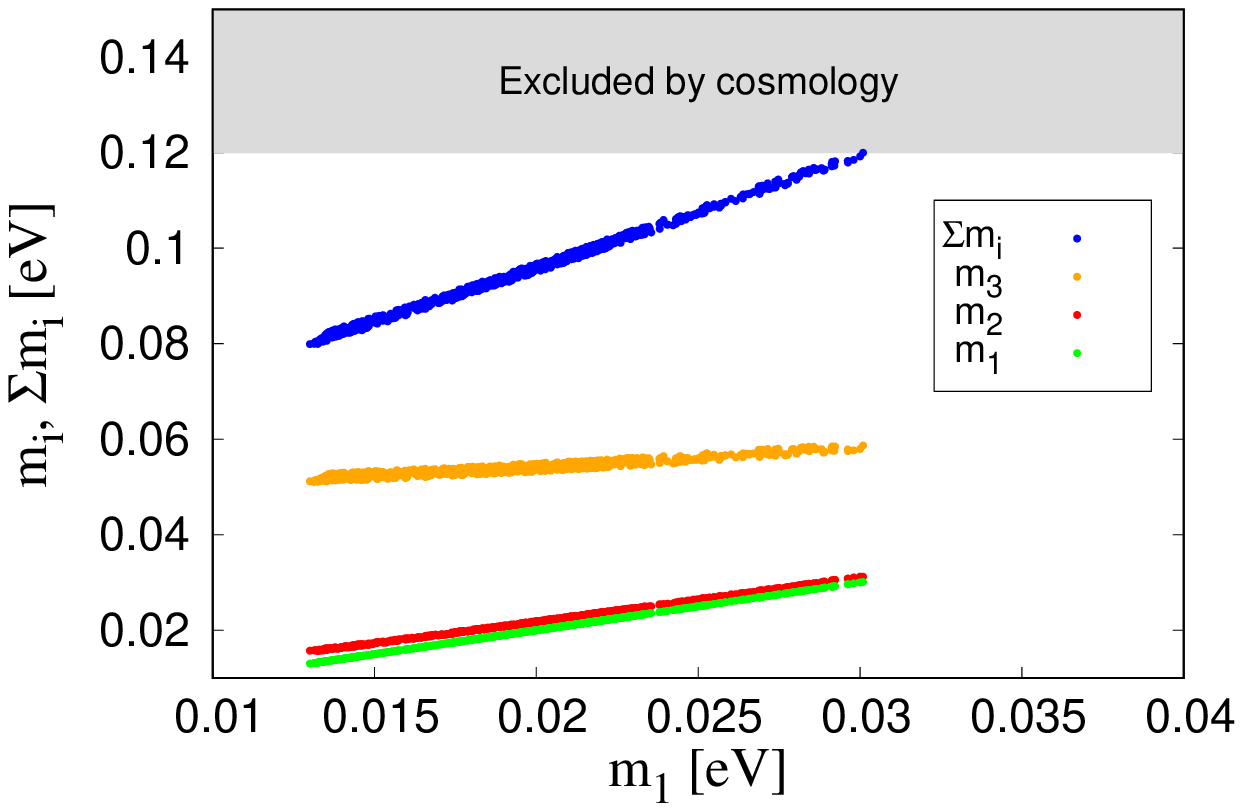}\quad %
\includegraphics[width=.44\textwidth]{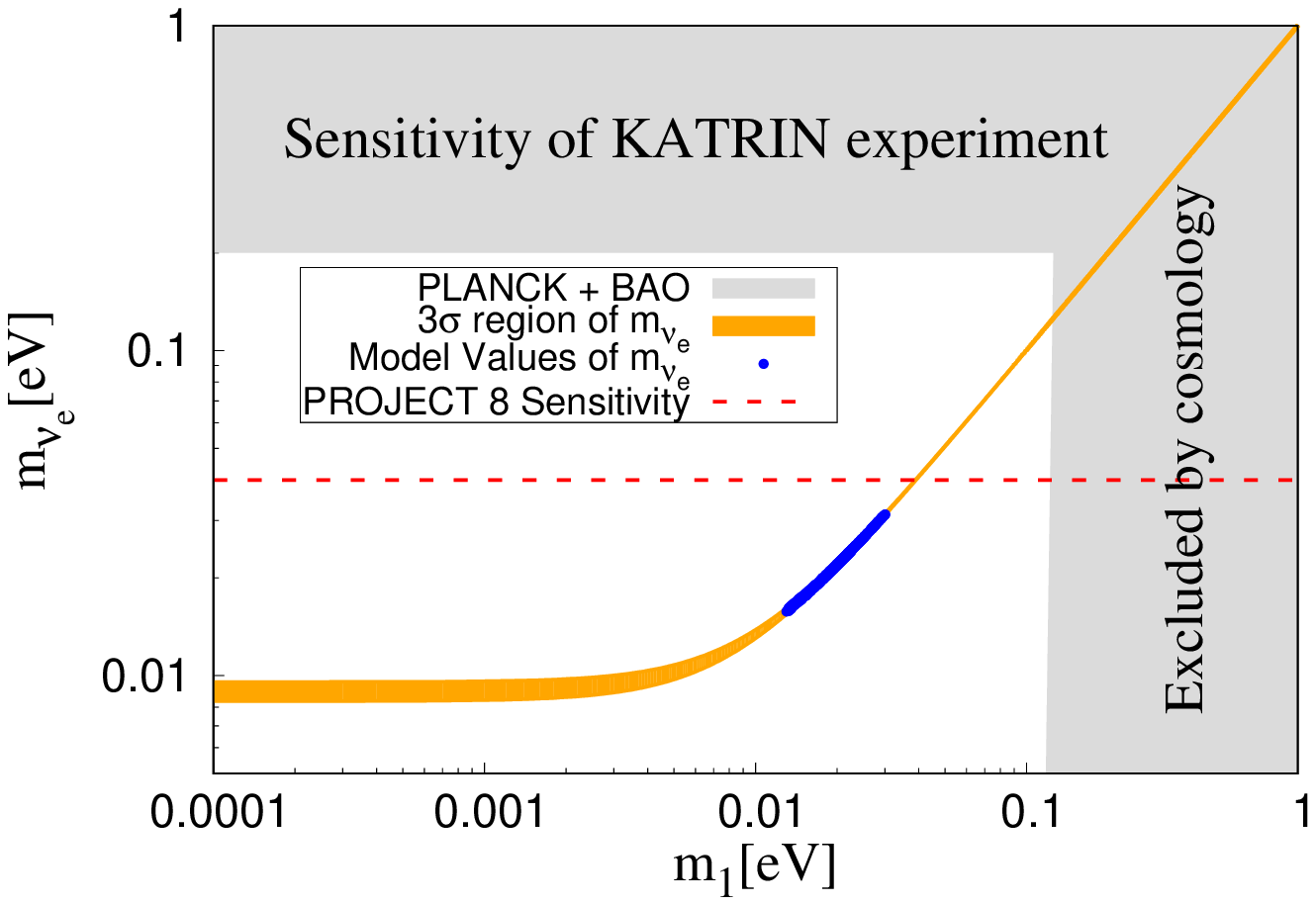}
\end{center}
\par
\vspace{0.5cm}
\caption{Left: Predictions for the absolute neutrino masses $m_{1}$ (green),
$m_{2}$ (red), $m_{3}$ (orange) and their sum $\sum m_{i}$ (blue) as a
function of the lightest neutrino mass $m_{1}$. The horizontal gray band
represent the upper limit on $\sum m_{i}$ provided by Planck+BAO data.
Right: The effective electron neutrino mass $m_{\protect\nu _{e}}$ as a
function of $m_{1}$. The vertical (horizontal) gray region is disfavored by
Planck+BAO (KATRIN ) data.}
\label{05}
\end{figure}
We start by substituting the elements of the mixing matrix and the masses
defined in the above observables by their expressions given in Eqs. (\ref%
{3-11}) and (\ref{3-12}) respectively. Hence, this shows the dependence of
these observables on our model parameters $a$, $b$, $\Gamma _{1}$\ and $\phi
_{a}$\ as well as the parameters involved in the trimaximal mixing matrix (%
\ref{3-11}). Then, we present our predictions using scatter plots. At first,
we show in the left panel of Fig. \ref{05}\ the correlations of the three
neutrino masses $m_{i=1,2,3}$ and their sum $m_{sum}$ versus the lightest
neutrino mass $m_{1}$ where we find%
\begin{eqnarray}
0.01300 &\lesssim &m_{1}\left[ \mathrm{eV}\right] \lesssim 0.03009\quad
,\quad 0.01561\lesssim m_{2}\left[ \mathrm{eV}\right] \lesssim 0.03124
\notag \\
0.05103 &\lesssim &m_{3}\left[ \mathrm{eV}\right] \lesssim 0.05867\quad
,\quad 0.07990\lesssim m_{sum}\left[ \mathrm{eV}\right] \lesssim 0.11997.
\end{eqnarray}%
From the interval of $m_{1}$, we take the lightest (largest) value and we
replace $m_{2}$ and $m_{3}$ by $\sqrt{m_{1}^{2}+\Delta m_{21}^{2}}$ and $%
\sqrt{m_{1}^{2}+\Delta m_{31}^{2}}$ respectively, we find that the sum of
neutrino masses\ in the normal mass hierarchy---using the best fit values of
$\Delta m_{21}^{2}$ and $\Delta m_{31}^{2}$ given in \textrm{\cite{R67}}%
---requires $m_{sum}\gtrsim 0.039\mathrm{eV}$ ($m_{sum}\gtrsim 0.09\mathrm{eV%
}$). While the constraint on $m_{sum}$ corresponding to the lightest $m_{1}$
is far from any current experiment, the upper bound on $m_{sum}$
corresponding to the largest $m_{1}$ may be achieved in the upcoming
experiments such as CORE+BAO targeting a bound on $\sum m_{i}$\ around $0.062%
\mathrm{eV}$ \textrm{\cite{R69}}.\textrm{\ }In the right panel of Fig. \ref%
{05}, we show the correlation between the effective mass of the electron
neutrino $m_{\nu _{e}}$ and $m_{1}$ where the horizontal gray band indicates
the expected sensitivity of $m_{\nu _{e}}$ from the KATRIN collaboration%
\textrm{\ \cite{R70,R71}}. We find that $m_{\nu _{e}}$ varies in the
following range%
\begin{equation}
0.01574\lesssim m_{\nu _{e}}\left[ \mathrm{eV}\right] \lesssim 0.03133.
\end{equation}%
Clearly, the values in this interval are very small when compared with the
forthcoming $\beta $-decay experiment sensitivities such as KATRIN \textrm{%
\cite{R70,R71}}, HOLMES \textrm{\cite{R72}}, and Project 8 \textrm{\cite{R73}%
} which will investigate $m_{\nu _{e}}$ at $0.2$\textrm{eV, }$0.1$\textrm{eV
and }$0.04$\textrm{eV} respectively. If none of these experiments would
measure $m_{\nu _{e}}$, our predicted values could be probed by future
experiments aiming to reach improved sensitivities around $0.01\mathrm{eV}$.
\begin{figure}[th]
\begin{center}
\hspace{2.5em} \includegraphics[width=.54\textwidth]{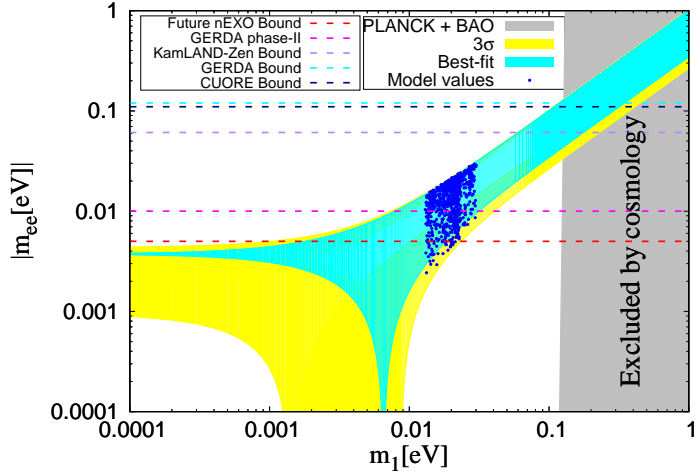}\quad
\end{center}
\par
\vspace{0.5cm}
\caption{The effective Majorana mass $\left\vert m_{ee}\right\vert $ as a
function of the lightest neutrino mass $m_{1}$. The vertical gray region
indicates the upper limit on the sum of the three light neutrino masses from
Planck+BAO data.}
\label{06}
\end{figure}
Now, let us explore the effective Majorana neutrino mass parameter of
neutrinoless double beta decay $\left\vert m_{ee}\right\vert $. A positive
signal of $0\nu \beta \beta $ would assert the Majorana nature of neutrinos
as well as provide a measure of the absolute neutrino mass scale.\textrm{\ }%
There are many ongoing and upcoming\ experiments around the world setting as
their purpose the detection of this process, where the present bounds on $%
\left\vert m_{ee}\right\vert $\ come from the KamLAND-Zen \textrm{\cite{R74}}%
, CUORE \textrm{\cite{R75}} and GERDA \textrm{\cite{R76}} experiments
corresponding to $\left\vert m_{ee}\right\vert <\left( 0.061-0.165\right)
\mathrm{eV}$, $\left\vert m_{ee}\right\vert <\left( 0.11-0.5\right) \mathrm{%
eV}$\ and $\left\vert m_{ee}\right\vert <\left( 0.15-0.33\right) \mathrm{eV}$%
\ respectively.\emph{\ }In Fig. \ref{06}, we show the correlation between $%
\left\vert m_{ee}\right\vert $ and the lightest neutrino mass $m_{1}$ where
we use the known $3\sigma $ ranges of the oscillation parameters\ while we
allow all phases to vary between $0$ and $2\pi $. Thus, the interval of $%
\left\vert m_{ee}\right\vert $ is given by%
\begin{equation}
0.00316\lesssim \left\vert m_{ee}\left[ \mathrm{eV}\right] \right\vert
\lesssim 0.02949.
\end{equation}%
As a result, our predicted region is far from the current sensitivities as
can be seen from the horizontal dashed lines in Fig. \ref{06} displaying the
bounds on $\left\vert m_{ee}\right\vert $ from some of the ongoing $0\nu
\beta \beta $ decay experiments. On the other hand, the anticipated
sensitivities of the next-generation experiments such as GERDA Phase II ($%
\left\vert m_{ee}\right\vert \sim \left( 0.01-0.02\right) \mathrm{eV}$)
\textrm{\cite{R77}} and nEXO ($\left\vert m_{ee}\right\vert \sim 0.005%
\mathrm{eV}$) \textrm{\cite{R78}}\ will cover our model predictions on $%
\left\vert m_{ee}\right\vert $.

On a different note,\ the expression for Jarlskog rephasing quantity $J_{CP}=%
\func{Im}(U_{e1}U_{\mu 1}^{\ast }U_{\mu 2}U_{e2}^{\ast })$\ which is a
measure of $CP$ violation, is given in terms of the trimaximal parameters\ $%
\sigma $\ and $\theta $\ as follows%
\begin{equation}
J_{CP}=\frac{\sin 2\theta \sin \sigma }{6\sqrt{3}}.  \label{JCP}
\end{equation}%
On the other hand, by matching the expressions of the rephasing invariant $%
J_{CP}$\ in the standard parametrization of the PMNS matrix and in the
trimaximal matrix defined in Eq. (\ref{3-11}), we derive the relation
between the Dirac $CP$ phase $\delta _{CP}$\ and the arbitrary phase $\sigma
$\ given as $\sin \sigma =\sin 2\theta _{23}\sin \delta _{CP}$. Moreover,
recall that the trimaximal mixing approach used in this model restricts the
atmospheric angle $\theta _{23}$\ around its maximal value (but not exactly
maximal $\theta _{23}\neq 45$) while the $\delta _{CP}$\ phase falls in the
close vicinity of $\delta _{CP}\simeq 0.5\pi $\ and $\delta _{CP}\simeq
-0.5\pi $.
\begin{figure}[th]
\begin{center}
\hspace{2.5em} \includegraphics[width=.44\textwidth]{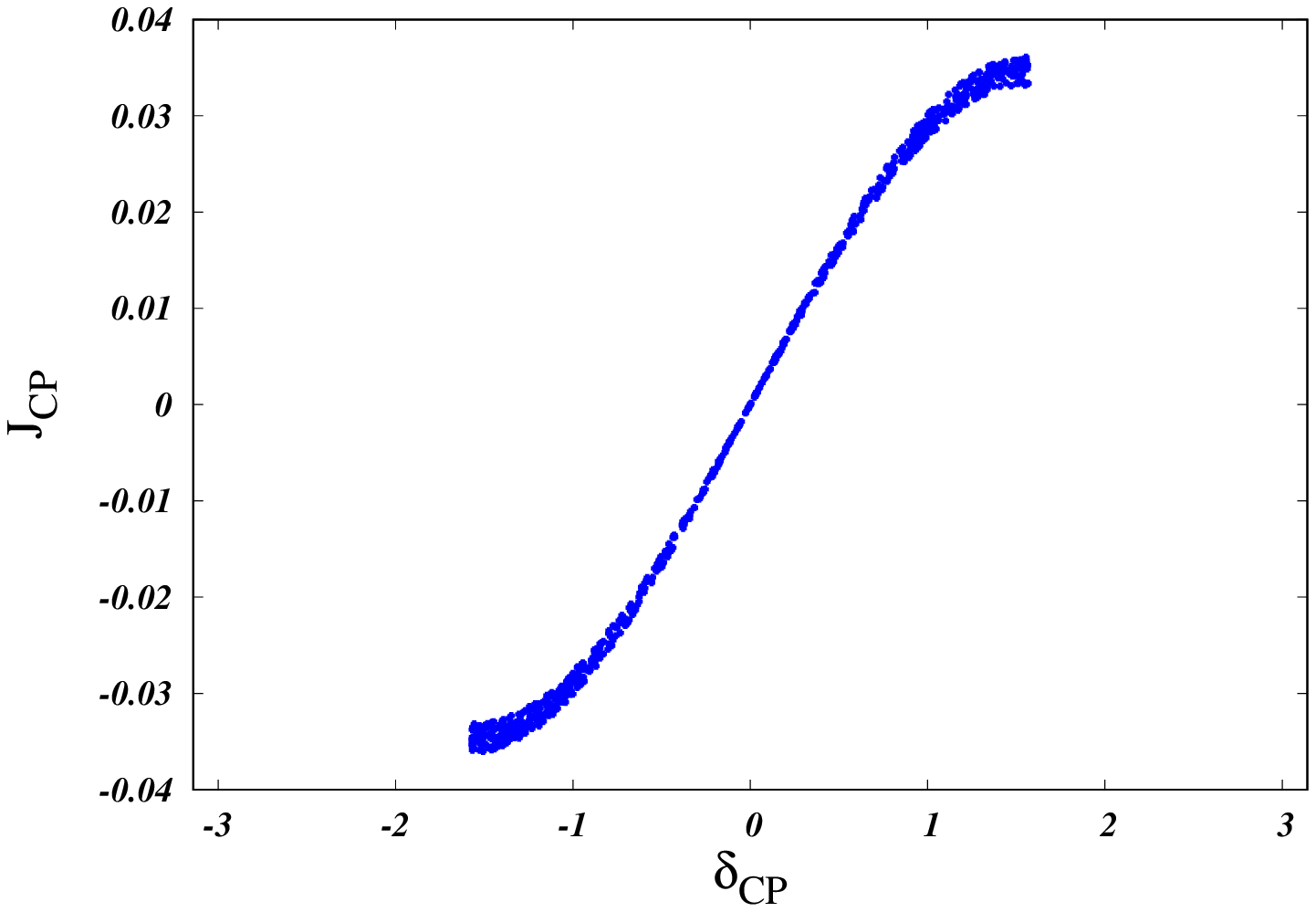}\quad %
\includegraphics[width=.44\textwidth]{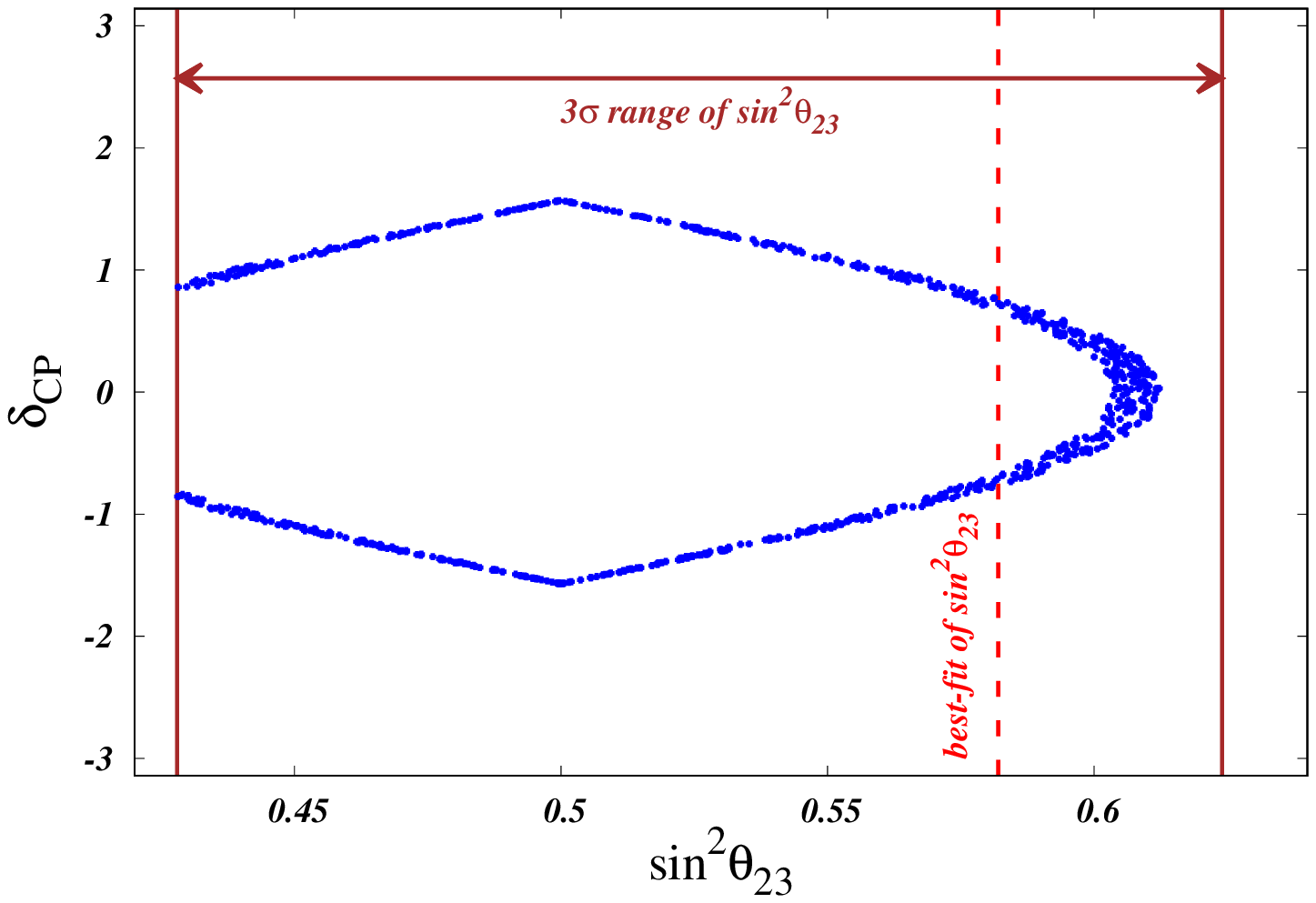}
\end{center}
\par
\vspace{0.5cm}
\caption{Left : Correlation between the Jarlskog invariant $J_{CP}$ and the
CP-violating phase $\protect\delta _{CP}$. Right: Correlation between $%
\protect\delta _{CP}$ and the atmospheric angle $\sin ^{2}\protect\theta %
_{23}$.}
\label{07}
\end{figure}
In this regards, we show in left panel of Fig. \ref{07} the correlation
between $J_{CP}$\ and the $\delta _{CP}$\ phase where we used the equation
relating $\delta _{CP}$\ with $\sigma $\ as an input, and since we have the
same parameter scan as before, the obtained range for the arbitrary phase $%
\sigma $ is as in Eq. (\ref{sig}) while for $\delta _{CP}$ we find%
\begin{equation}
-0.499\pi \lesssim \delta _{CP}\left[ \mathrm{rad}\right] \lesssim 0.498\pi .
\label{delt}
\end{equation}%
\newline
Based on Eq. (\ref{JCP}) and the obtained ranges of $\theta $\ and $\sigma $%
; it is straightforward to verify that $\sin 2\theta \neq 0$\ and $\sin
\sigma \neq 0$, subsequently leading to a non vanishing $J_{CP}\neq 0$. On
the other hand, it is clear from Eq. (\ref{mix}) that the maximal value of
the atmospheric angle ($\sin ^{2}\theta _{23}=1/2$) is excluded in our
model. Moreover, since\ the $CP$ phase is correlated significantly with $%
\sin ^{2}\theta _{23}$\ than the other mixing angles, we show in the right
panel of Fig. \ref{07} the predicted regions of $\delta _{CP}$\ versus $\sin
^{2}\theta _{23}$\ at $3\sigma $. As can be seen, our model allows $\sin
^{2}\theta _{23}$\ to vary randomly in the interval $[0.428\rightarrow
0.612] $\ where the corresponding region of $\delta _{CP}$\ is as in Eq. (%
\ref{delt}).

\section{Dark Matter candidates}

Before we provide the possible DM candidates in the present model, let us
discuss the breaking pattern of the $\emph{G}_{f}$ flavor group. First,
recall that our model involves three scalar fields that acquire VEVs; the
usual $SU(2)_{L}$ Higgs doublet $H$\ and two flavon fields $\digamma $\ and $%
\chi $. The Higgs doublet $H$\ transforms trivially under $\emph{G}_{f}$ and
thus, it only contributes to the EW symmetry breaking. On the other hand,
the flavons $\digamma $\ and $\chi $\ transform respectively as a\ doublet
and a trivial singlet under\textrm{\ }$\mathbb{D}_{4}$. Therefore, only the
nontrivial VEV of $\digamma $\ is responsible for breaking the\textrm{\ }$%
\mathbb{D}_{4}$\textrm{\ }symmetry down to one of its subgroups\footnote{%
Notice that since the flavon field $\chi $\ carries the charges $\omega $\
and $\eta ^{4}$\ under $Z_{3}$\ and $Z_{5}$\ respectively, these groups are
spontaneously broken to the identity.}\textrm{\ }$G_{r}\subset \mathbb{D}%
_{4} $. Moreover, since these three scalar fields are chosen to be even
under the additional $Z_{2}$ symmetry, this latter remains unbroken.\textbf{%
\ }To determine the remnant $G_{r}$\ symmetry that survives the $\mathbb{D}%
_{4}$\ breaking, we recall that $\mathbb{D}_{4}$\ is isomorphic to the
semidirect product $Z_{4}\rtimes Z_{2}^{\prime }$ and has two generators $S$%
\ and $T$\ where $S$ generates $Z_{4}$\ and $T$ generates $Z_{2}^{\prime }$\
symmetries\ satisfying the relations $S^{4}=T^{2}=Id$ and $STS=T$.\textrm{\ }%
Recall also that the VEV alignment of the flavon $\digamma $---$\left\langle
\digamma \right\rangle =\upsilon _{\digamma }(1,1)$---is chosen to reproduce
the observed neutrino masses and mixing.\textrm{\ }This specific VEV
direction breaks\textrm{\ }$\mathbb{D}_{4}$\textrm{\ }down to $Z_{2}^{\prime
}$ with broken part given by the $Z_{4}$\ group. This means that the VEV
structure of the flavon $\digamma $\ preserves the generator $T$\ while
changes the generator $S$; we have\footnote{%
See the matrix representation of the\textrm{\ }$\mathbb{D}_{4}$\textrm{\ }%
generators in the Appendix.}%
\begin{equation}
T\left\langle \digamma \right\rangle =\left\langle \digamma \right\rangle
\quad ,\quad S\left\langle \digamma \right\rangle \neq \left\langle \digamma
\right\rangle .
\end{equation}%
Therefore, the spontaneous breaking of the full\ discrete flavor symmetry $%
\emph{G}_{f}$\ is given by%
\begin{equation}
\mathbb{D}_{4}\times Z_{3}\times Z_{5}\times Z_{2}\overset{\left\langle
\digamma \right\rangle ,\left\langle \chi \right\rangle }{\longrightarrow }%
Z_{2}^{\prime }\times Z_{2}.
\end{equation}%
Now, we are in position to discuss the stabilization of the DM candidates by
the remnant $Z_{2}^{\prime }\times Z_{2}$\ symmetry. For this purpose, let
us first briefly comment these reflection symmetries individually. On the
one hand, for the residual $Z_{2}^{\prime }$\ symmetry, it is useful to
study the decomposition of $\mathbb{D}_{4}$ irreducible representations into
those of its subgroup $Z_{2}^{\prime }$. The latter has two singlet
representations $1_{+}$\ (trivial) and $1_{-}$, and from the characters of
the $\mathbb{D}_{4}$ group (see Table \ref{st} in the Appendix), it is easy
to check that the singlet representations $1_{+,+}$\ and $1_{+,-}$\ of $%
\mathbb{D}_{4}$ correspond to $1_{+}$\ of $Z_{2}^{\prime }$, while $1_{-,+}$%
\ and $1_{--}$\ of $\mathbb{D}_{4}$ correspond to $1_{-}$\ of $Z_{2}^{\prime
}$. For the $\mathbb{D}_{4}$ doublet $2$, it decomposes into $Z_{2}^{\prime
} $\ representations as $2=1_{+}+1_{-}$\ where the first component of $2$\
is associated to $1_{+}$\ while the second one is associated to $1_{-}$.
Therefore, the particles running in the loop transform under the $%
Z_{2}^{\prime }$ symmetry as%
\begin{equation}
\begin{array}{ccccccccccc}
N_{1} & \rightarrow & N_{1} & , & N_{2} & \rightarrow & N_{2} & , & N_{3} &
\rightarrow & -N_{3} \\
&  & \rho & \rightarrow & \rho & , & \phi & \rightarrow & \phi &  &
\end{array}
\label{z2}
\end{equation}%
From these transformations, it is easy to notice that this remaining $%
Z_{2}^{\prime }$ symmetry is not sufficient for DM stabilization. For
example, in the case where $\rho $\ is the DM candidate, the couplings $\bar{%
L}_{i}l_{i}^{R}\rho $\ with $i=e$,$\mu ,\tau $ are protected\ by $%
Z_{2}^{\prime }$\ after symmetry breaking, therefore these couplings lead to
the DM decay $\rho \rightarrow \bar{L}_{i}l_{i}^{R}$.

On the other hand, the particles\ in (\ref{z2}) are odd under the extra
symmetry $Z_{2}$\ whilst all SM particles are even under it. This clearly
shows that this extra symmetry stabilizes these potential DM particles
against decay into SM ones. Therefore, the DM candidate in our model is the
lightest among the fermionic right-handed neutrinos $N_{k}$\ and the scalars
$\rho $\ and $\phi $. Moreover, since the full residual flavor symmetry in
the neutrino sector is $Z_{2}^{\prime }\times Z_{2}$\ group, there might be
processes allowed by $Z_{2}$\ but forbidden by $Z_{2}^{\prime }$. Thus, it
is important to verify the invariance of the various DM processes under $%
Z_{2}$ as well as $Z_{2}^{\prime }$. Here, we will discuss briefly the
possible DM candidates while a thorough calculation of their properties such
as annihilation cross section, lifetime and relic abundance is beyond the
purpose of the present work. \newline
As mentioned above, all the particles running in the loop diagram of Fig. %
\ref{02} are potential DM candidates. In the following, we discuss\emph{\ }%
two possibilities:

\textbf{Case I: Fermionic dark matter candidate}

In order to facilitate the engineering of neutrino masses and mixing, we
have considered the case where $M_{3}\simeq M_{2}\simeq M_{1}$. We assume
here for simplicity that $N_{3}$\ is the only fermionic DM candidate. In
this scenario, a pair of $N_{3}$ can annihilate into SM particle pair $%
N_{3}N_{3}\rightarrow l^{+}l^{-}$($\nu \bar{\nu}$)\ through t-channel
diagrams mediated by the components\ of scalar doublets $\rho $\ and $\phi $
given in (\ref{3-3}). Moreover, it is well known that the dark matter
abundance depends not only on the annihilation cross section, but for
quasi-degenerate states the co-annihilation cross section\ becomes
important. In the present case, since the neutrino masses $M_{1}$\ and $%
M_{2} $\ are close to $M_{3}$\ (which implies the relative mass difference $%
\Delta _{i}=\frac{(M_{i}-M_{3})}{M_{3}}\ll 1$), then the co-annihilation
processes of $N_{3}N_{1}$\ and $N_{3}N_{2}$\ into charged leptons $%
l^{+}l^{-} $ and neutrino $\nu \bar{\nu}$ dominate over the annihilation
processes. On the other hand, since the right handed neutrino $N_{3}$\
transforms as an $SU(2)_{L}$\ singlet, it has no tree level couplings to the
Higgs boson or the $Z$ boson.\textrm{\ }However, due to the Majorana nature
of the DM particle, the spin independent (SI) scattering cross section of $%
N_{3}$\ on nucleons takes place via the effective coupling $y_{hN_{3}N_{3}}$%
that induces the one-loop effective coupling between DM and the Higgs boson
\textrm{\cite{X2,X3}}. \newline
The right handed neutrino as a suitable DM candidate have been studied in
several radiative neutrino models with inert Higgs doublet, see for example\
\textrm{\cite{R34,X4}}. In particular, as a result of the analysis performed
in reference\footnote{%
The involved processes for annihilation, co-annihilation as well as SI
scattering of DM in this study are roughly similar to the ones involved in
our model.}\textrm{\ \cite{R34}}, the RH neutrino is a promising DM
candidate in the mass range $10<m_{DM}$\ (GeV)$<700$.\ This region is in
agreement with the relic abundance measured by WMAP\textrm{\ \cite{X6} }and
Planck\textrm{\ \cite{X7} }collaborations and with a SI direct detection
cross section below the upper limits given by Lux\textrm{\ \cite{X8} }and
Xenon1T\textrm{\ \cite{X9} }experiments.

\textbf{Case II: Scalar dark matter candidate}

We assume that the DM candidate is one of the neutral components of the
inert scalar doublet $\rho $\ ($\rho _{1}$\ or $\rho _{2}$) and it is
lighter than the flavon fields $\chi $\ and $\digamma $\ and the members of
the scalar triplet $T$. Therefore, the relic abundance of DM can only be
determined by its annihilation/co-annihilation to SM particles. Moreover,
since $\rho $\ is an EW doublet, it is well known that such processes are
mediated by SM Higgs\ and gauge bosons. The relevant invariant terms
involving the inert doublet $\rho $\ in the scalar potential can be written
as%
\begin{eqnarray}
\mathcal{V}\left( \rho \right) &\supset &\mu _{\rho }^{2}\left\vert \rho
\right\vert ^{2}+2\lambda _{1}\left( \rho ^{\dagger }\rho \right)
^{2}+2\lambda _{2}(\rho ^{\dagger }\digamma _{1})(\rho \digamma
_{1}^{\dagger })+2\lambda _{3}(\rho ^{\dagger }\rho )(\digamma \digamma
^{\dagger })  \notag \\
&&+\lambda _{4}(H^{\dagger }H)(\rho ^{\dagger }\rho )+\lambda
_{5}(H^{\dagger }\rho )(\rho ^{\dagger }H)+\lambda _{6}(\rho ^{\dagger }\chi
)(\rho \chi ^{\dagger })+\lambda _{7}(\rho ^{\dagger }\rho )(\chi ^{\dagger
}\chi )  \label{V} \\
&&+\lambda _{8}(\rho ^{\dagger }\rho )Tr\left( T^{\dagger }T\right) +\lambda
_{9}\rho ^{\dagger }TT^{\dagger }\rho  \notag
\end{eqnarray}%
\textrm{\ }As mentioned above, the SM Higgs doublet $H$, the triplet Higgs $%
T $\ and the flavon singlets $\digamma $\ and $\chi $\ can develop VEVs
while the inert doublet $\rho $\ does not develop a VEV. This can be
parameterized as
\begin{eqnarray}
\left\langle H\right\rangle &=&\frac{1}{\sqrt{2}}\left(
\begin{array}{c}
0 \\
\upsilon _{H}+h%
\end{array}%
\right) ~,~\rho =\frac{1}{\sqrt{2}}\left(
\begin{array}{c}
\rho _{1}+i\rho _{2} \\
\sqrt{2}\rho ^{-}%
\end{array}%
\right)  \notag \\
\left\langle T\right\rangle &=&\frac{1}{\sqrt{2}}\left(
\begin{array}{cc}
0 & 0 \\
\upsilon _{T}+T & 0%
\end{array}%
\right) ~,~\left\langle \digamma \right\rangle =\frac{1}{\sqrt{2}}\left(
\upsilon _{\digamma }+\digamma \right) ~,~\left\langle \chi \right\rangle =%
\frac{1}{\sqrt{2}}\left( \upsilon _{\chi }+\chi \right)
\end{eqnarray}%
where we have omitted the pseudo scalars for all the fields except for $\rho
$. From this parametrization, we find the masses of the physical states $%
\rho _{1}$, $\rho _{2}$\ and $\rho ^{\pm }$\ in terms of parameters of the
potential $\mathcal{V}\left( \rho \right) $\ as%
\begin{eqnarray}
M_{\rho _{1},\rho _{2}}^{2} &=&\mu _{\rho }^{2}+\left( \lambda _{2}+\lambda
_{3}\right) \upsilon _{\digamma }^{2}+\frac{\lambda _{4}}{2}\upsilon
_{H}^{2}+\frac{1}{2}\left( \lambda _{6}+\lambda _{7}\right) \upsilon _{\chi
}^{2}+\frac{\lambda _{8}}{2}\upsilon _{T}^{2}  \notag \\
M_{\rho ^{\pm }}^{2} &=&\mu _{\rho }^{2}+\left( \lambda _{2}+\lambda
_{3}\right) \upsilon _{\digamma }^{2}+\frac{1}{2}\left( \lambda _{4}+\lambda
_{5}\right) \upsilon _{H}^{2}+\frac{1}{2}\left( \lambda _{6}+\lambda
_{7}\right) \upsilon _{\chi }^{2}+\frac{1}{2}\left( \lambda _{8}+\lambda
_{9}\right) \upsilon _{T}^{2}
\end{eqnarray}%
As a result, we obtain a mass degeneracy between $\rho _{1}$\ and $\rho _{2}$%
\ and thus we cannot distinguish between them in the present model. This
degeneracy is due to the vanishing of the coupling $\left\{ \lambda _{\rho
}(H\rho )^{2}+h.c.\right\} $ which is---as in the usual inert doublet model
(IDM) \cite{A11}---responsible for the mass splitting of the neutral
components $\rho _{1}$\ and $\rho _{2}$\ of the inert doublet $\rho $\ after
EW symmetry breaking. The $\lambda _{\rho }$ term is actually invariant
under all the symmetries of the model; however, due to the\textrm{\ }$%
\mathbb{D}_{4}$\textrm{\ }structure of $\rho $\ given in (\ref{3-4}) the
tensor product\footnote{%
The hat in $\hat{\rho}$\ denotes a $\mathbb{D}_{4}$ doublet while $\rho $
without the hat character denotes the $SU(2)_{L}$ inert doublet; see Eq. (%
\ref{3-4}).} $\left. \hat{\rho}\otimes \hat{\rho}\right\vert _{1_{++}}$\
vanishes according to Eq. (\ref{A2}) in the appendix. Therefore, the\textrm{%
\ }$\mathbb{D}_{4}$\textrm{\ }structure of $\rho $\ preserves the mass
degeneracy between $\rho _{1}$\ and $\rho _{2}$. Reasoning from this outcome
and the fact that the couplings of DM particles with gauge bosons relate
directly to the cross section for scattering off a nucleus, the DM-quark
inelastic scattering $\rho _{1}q\rightarrow \rho _{2}q$\ can be described by
the unsuppressed vertex coupling $Z-\rho _{1}-\rho _{2}$ whose size is fixed
by the EW gauge coupling. Such DM inelastic scattering scenario predicts a
large cross section in the direct detection experiments and has already
excluded by the current results \cite{A12,A13,A14}.\newline
Regardless of the smallness of the mass splitting required to fit the
experimental data, one way to allow a mass splitting between the neutral
component of the inert scalar $\rho $\ in the present model is by modifying
the $\mathbb{D}_{4}$ irreducible representation of $\rho $\ where if it is
assigned to one of the $\mathbb{D}_{4}$ singlets instead of the $\mathbb{D}%
_{4}$ doublet, the $\lambda _{\rho }$\ term will be allowed and then, the
mass splitting between $\rho _{1}$\ and $\rho _{2}$\ will depend on the size
of this quartic coupling $\lambda _{\rho }$. However, assigning the\textrm{\
}$\mathbb{D}_{4}$\textrm{\ }doublet to $\rho $\ is required to make the
topology T4-2-i genuine which is our primary concern in the present study.
Notice by the way that the discussion on $\rho $\ holds as well for the case
of the scalar $\phi $ where by replacing $\rho $\ by $\phi $\ in Eq. (\ref{V}%
), we end up with the same conclusion for the masses of the chargeless
components of\textrm{\ }$\phi $\textrm{.}

\section{Conclusion}

In this work, we have proposed a radiative neutrino model based on topology
T4-2-i providing an explanation for the observed neutrino masses and mixing
as well as allowing for stable dark matter candidates. In order to avoid the
tree level Type I and Type II seesaw contributions always accompanying
topology T4-2-i, fitting the neutrino data as well as to ensure the
stability of DM candidates, we have extended the SM gauge symmetry with the $%
\emph{G}_{f}=\mathbb{D}_{4}\times Z_{3}\times Z_{5}\times Z_{2}$ flavor
group. For this purpose, besides promoting the fermion in topology T4-2-i to
singlet right-handed neutrinos $N_{k}$, we have added two flavon fields $%
\digamma $ and $\chi $ to guarantee the invariance of neutrino Yukawa
couplings and the preexisting vertex $\mu _{H}HT^{\dagger }H$\ connecting
two Higgs doublets with the scalar triplet $T$. Therefore, the neutrino
masses are radiatively generated at one-loop level while their mixing is
described by the well known TM$_{2}$ pattern due to $\emph{G}_{f}$ symmetry
breaking.\newline
We have performed our numerical study in the normal mass hierarchy case
where we have shown through several scatter plots the allowed ranges of our
model parameters $\left\{ \sigma ,\theta ,a,b,\phi _{a},\Gamma _{1}\right\} $%
, which we have used to predict the ranges of the $CP$ violating phase $%
\delta _{CP}$ as well as the non-oscillatory observables $m_{\nu _{e}}$, $%
\left\vert m_{ee}\right\vert $ and\textrm{\ }$m_{sum}$ that fit the
experimental values of the three mixing angles $\theta _{ij}$\ and the mass
square differences $\Delta m_{ij}^{2}$\ at $3\sigma $\ range.

Another matter considered in the present work is the dark matter candidates
given by one of the fields running in the loop. On the basis of our
considerations,\emph{\ }the DM candidates can be manifested by the neutral
components of the scalar doublets $\rho $\ and $\phi $, and the right-handed
neutrino $N_{3}$, the lightest of which can play the role of DM as they
carry odd\ charge under $Z_{2}$. We showed that the stability of DM is
guaranteed by the unbroken $Z_{2}$\ symmetry while the different DM
processes are controlled by the residual group $Z_{2}^{\prime }\times Z_{2}$%
\ after $G_{f}$\ symmetry breaking. In the case of the inert scalar $\rho $,
we found that there are no mass splitting between its neutral components%
\emph{\ }where in the scenario of DM inelastic scattering predicts a large
cross section in the direct detection experiments and has already excluded
by the current data. On the other hand, in the case of Majorana DM, $N_{3}$
is a suitable candidate because the relevant processes for annihilation,
co-annihilation as well as SI scattering of DM in our study are roughly
similar to the ones involved in radiative models with right-handed Majorana
DM.\emph{\ }We should mention however that a thorough study of the latter
case requires performing further studies by analysing two particular
experimental constraints; the observed DM relic density and the cross
section for direct detection of DM scattering off nucleon. This analysis,
however, goes beyond the scope of this paper.

\section{Appendix: Dihedral $\mathbb{D}_{4}$ group}

In this appendix, we briefly review the basic features of the dihedral group
$\mathbb{D}_{4}$\ as well as the decomposition of its representations into
those of the $Z_{2}^{\prime }$ subgroup. Recall first that the discrete
group $\mathbb{D}_{4}$\ is generated by the two elements $S$\ and $T$\ which
fulfill the relations $S^{4}=T^{2}=Id$\ and $STS=T$. It has five irreducible
representations; four singlets $1_{+,+},1_{+,-},1_{-,+}$\ and $1_{-,-}$, and
one doublets $2$\ where the indices in the representations refer to their
characters under $S$\ and $T$\ as in the following table
\begin{table}[h]
\centering
\begin{tabular}{|c|c|c|c|c|c|}
\hline
$\mathrm{\chi }_{i,j}$ \  & $\mathrm{\chi }_{1_{+,+}}$ & $\mathrm{\chi }%
_{1_{+,-}}$ & $\mathrm{\chi }_{1_{-,+}}$ & $\mathrm{\chi }_{1_{-,-}}$ & $%
\mathrm{\chi }_{2_{0,0}}$ \\ \hline
$Id$ & $+1$ & $+1$ & $+1$ & $+1$ & $2$ \\ \hline
$S$ & $+1$ & $-1$ & $+1$ & $-1$ & $0$ \\ \hline
$T$ & $+1$ & $+1$ & $-1$ & $-1$ & $0$ \\ \hline
\end{tabular}%
\caption{Character table of the dihedral group $\mathbb{D}_{4}$.}
\label{st}
\end{table}
The generators $S$\ and $T$\ of the two-dimensional representations can be
expressed by the following $2\times 2$\ matrices%
\begin{equation}
S=\left(
\begin{array}{cc}
i & 0 \\
0 & -i%
\end{array}%
\right) \quad ,\quad T=\left(
\begin{array}{cc}
0 & 1 \\
1 & 0%
\end{array}%
\right) .
\end{equation}%
Now, we consider tensor products of $\mathbb{D}_{4}$ irreducible
representations.\textrm{\ }The tensor product of two doublets $%
2_{x}=(x_{1},x_{2})^{T}$\ and $2_{y}=(y_{1},y_{2})^{T}$\ is decomposed into
a sum of $\mathbb{D}_{4}$\ singlet representations as\textrm{\ }$2_{x}\times
2_{y}=1_{+,+}+1_{+,-}+1_{-,+}+1_{-,-}$, where%
\begin{equation}
\begin{array}{ccc}
1_{+,+}=x_{1}y_{2}+x_{2}y_{1} & , & 1_{+,-}=x_{1}y_{1}+x_{2}y_{2} \\
1_{-,+}=x_{1}y_{2}-x_{2}y_{1} & , & 1_{-,-}=x_{1}y_{1}-x_{2}y_{2}%
\end{array}
\label{A2}
\end{equation}%
whereas the product between two singlets is as follows%
\begin{equation}
1_{i,j}\times 1_{k,l}=1_{ik,jl}\text{ \ \ with \ }i,j,k,l=\pm \text{.}
\end{equation}%
Finally,\textrm{\ }since the experimental data on neutrino masses and mixing
angles require the breaking of $\mathbb{D}_{4}$\ down to its remnant $%
Z_{2}^{\prime }$\ subgroup, so we restrict our discussion only to the
breaking pattern $\mathbb{D}_{4}\overset{\left\langle \digamma \right\rangle
}{\longrightarrow }Z_{2}^{\prime }$.\textrm{\ }After this stage of breaking,
it is obvious that the matter and scalar fields in our model will be charged
under the unbroken discrete symmetry $Z_{2}^{\prime }$. Accordingly, we
summarize in the following table the decompositions of $\mathbb{D}_{4}$\
irreducible representations\ into those of the residual group $Z_{2}^{\prime
}$\ subgroup%
\begin{equation}
\begin{tabular}{lll}
\hline\hline
Fields under $\mathbb{D}_{4}$ Irreps. &  & Decomposition into $Z_{2}^{\prime
}$ Irreps. \\ \hline\hline
$\ \ \left(
\begin{array}{c}
\alpha _{1} \\
\alpha _{2}%
\end{array}%
\right) \sim 2$ & $\rightarrow $ & $\ \ \ \ \ \left.
\begin{array}{c}
\alpha _{1}\sim +1 \\
\alpha _{2}\sim -1%
\end{array}%
\right. $ \\ \hline
$\beta _{1}\sim 1_{+,+},\beta _{2}\sim 1_{+,-}$ & $\rightarrow $ & \ $\ \ \
\ \ \beta _{1},\beta _{2}\sim +1$ \\ \hline
$\gamma _{1}\sim 1_{-,+},\gamma _{2}\sim 1_{-,-}$ & $\rightarrow $ & \ $\ \
\ \ \ \ \gamma _{1},\gamma _{2}\sim -1$ \\ \hline
\end{tabular}
\label{dec}
\end{equation}%
where $\alpha _{i}$, $\beta _{i}$\ and $\gamma _{i}$ can be any fermionic or
bosonic field. For more details on the $\mathbb{D}_{4}$ Dihedral group see
for instance Ref. \cite{R79}.

\end{document}